\documentclass[11pt,twocolumn]{article}

\usepackage[margin=0.75in]{geometry}
\usepackage{graphicx}
\usepackage{amsmath,amsfonts,amssymb}
\usepackage[normalem]{ulem}
\usepackage{multirow}
\usepackage{makecell}
\usepackage{xcolor}
\usepackage{authblk}
\usepackage[numbers,sort&compress]{natbib}
\usepackage{url}
\usepackage{placeins}
\usepackage{hyperref}

\hypersetup{colorlinks=true,linkcolor=black,citecolor=black,urlcolor=blue}

\definecolor{gold}{rgb}{0.83, 0.69, 0.22}
\definecolor{cp_blue}{rgb}{0.3,0.4,0.7}
\definecolor{mh_green}{rgb}{0.3,0.8,0.3}

\newcommand{\DMC}[1]{\textcolor{black}{#1}}

\newcommand{\dpm}{D2508}
\newcommand{\angstrom}{\mbox{\normalfont\AA}}

\title{The phase diagram of fluid hydrogen near melting from machine-learned interatomic potentials trained on quantum Monte Carlo data}

\author[1]{Shubhang Goswami}
\author[2]{Hongwei Niu}
\author[3]{Ilnur Saitov}
\author[4,7]{Mathieu Istas}
\author[5]{Yubo Yang}
\author[6]{Scott Jensen}
\author[4]{Markus Holzmann}
\author[3]{Carlo Pierleoni}
\author[1]{David Ceperley\thanks{Corresponding author: ceperley@illinois.edu}}

\affil[1]{University of Illinois Urbana-Champaign, Urbana, Illinois, 61801, USA}
\affil[2]{Inner Mongolia University of Technology, School of New Energy, Ordos, 017010, China}
\affil[3]{University of L'Aquila, Via Vetoio 10, I-67010 L'Aquila, Italy}
\affil[4]{Univ. of Grenoble Alpes, CNRS, LPMMC, 38000 Grenoble, France}
\affil[5]{Department of Physics and Astronomy, Hofstra University, Hempstead, NY, 11549, USA}
\affil[6]{University of Louisville, Louisville, Kentucky, 40292, USA}
\affil[7]{Univ. Grenoble Alpes, CNRS, Institut N\'eel, 38042 Grenoble, France}

\date{\today}

\begin{document}
\maketitle

\begin{abstract}
\DMC{Dissociation of molecular hydrogen increasing temperature or pressure may either occur gradually or abruptly, via a phase transition. Despite considerable experimental and theoretical efforts, the phase diagram of dense hydrogen at temperatures above solid melting is not precisely known. Here, we}
perform classical and quantum Path Integral molecular dynamics simulations of dense hydrogen using two machine-learned interatomic potentials both trained on quantum Monte Carlo energies and forces. 
The resulting phase diagram is constructed using state-of-the-art techniques.
Both models show a first-order phase transition from the molecular to the atomic liquid terminating in critical points 
whose locations were determined using finite-size scaling techniques. 
Triple points, located at the highest temperature of the melting curve, are found at the intersection of the liquid-liquid lines with the melting curve 
\DMC{and should lead to clear signatures in ramp compression experiments.}

\end{abstract}

\section{Introduction}

Hydrogen exhibits a rich and not fully resolved phase diagram. Characterizing its behavior over a broad range of temperatures and pressures is important for understanding the structure and evolution of astrophysical systems such as stars, gas giants, and dense interstellar matter. Ever since the prediction of metallization by Wigner and Huntington~\cite{wigner1935possibility}, the observation of pressure-induced metallization has been a central objective in high-pressure physics.
\DMC{In the solid phase, experiments discovered a sequence of structural phase transitions \cite{Mao1994} of the molecular crystal under static compression, with no clear signs of dissociation, whereas dynamic compression experiments at higher temperatures show metallic behavior in the fluid phase~\cite{ginzburg_what_1999,zaghoo2016evidence,celliers2018insulator,gregoryanz2020everything}. }

Fluid hydrogen exists in two phases: a molecular insulating phase and an atomic metallic phase. 
For pressures less than 150~GPa and temperatures below 2500~K fluid hydrogen exists primarily as an insulating molecular liquid. Upon increasing pressure or temperature the molecules dissociate turning the system into an atomic fluid. 
However, since there is no difference in symmetry between fluid molecular hydrogen and fluid atomic hydrogen, whether the change happens via a crossover or a phase transition has been a long-standing question~\cite{bonitz_first_2024,cheng2020evidence,karasiev2021liquid,tirelli2022high,istas2025liquid} since the work of Zeldovitch and Landau~\cite{Zeldovitch}.

The first evidence of a phase transition was provided by Density Functional Theory (DFT) based First-Principles Molecular Dynamics (FPMD) simulations in the NPT ensemble~\cite{scandolo2003liquid}. 
Since then, the first-order nature of the transition has been confirmed by many DFT-FPMD studies~\cite{morales2010evidence,lorenzen2010first,karasiev2021liquid,norman2017critical} (see ref~\cite{bonitz_first_2024} for an extended bibliography).
However, the location of the transition line is sensitive to the employed density functional as well as to nuclear quantum effects~\cite{pierleoni2016liquid}. 
Simulation methods based on Quantum Monte Carlo (QMC) which are expected to represent the potential energy surface of the protons more accurately, such as the Coupled Electron-Ion Monte Carlo method (CEIMC)~\cite{pierleoni2016liquid,pierleoni2018CPP, bonitz_first_2024} and the QMC-MD~\cite{Mazzola2018}, also supported the existence of a first-order transition. Although CEIMC and QMC-MD are significantly different methods and are based on different QMC wave functions, their  predictions about the LLPT transition line are in reasonable agreement.
Although hydrogen, with its simple electronic structure and the absence of core electrons, is a favorable system to apply First-Principles methods, studying phase transition and critical points at the ab-initio level is still quite challenging. Establishing the nature of the transition and investigating the occurrence of the critical point requires spanning a wide range of system sizes and simulation times. This investigation can be achieved today by employing Machine Learnt force fields trained on datasets of nuclear configurations with first-principle features (energy, forces and stress). In principle this preserves a first-principle accuracy in predicting new configurations but allows one to perform simulations of much larger systems for much longer times.
Subsequently, Istas et al.~\cite{istas2025liquid} investigated the LLPT with finite-size scaling methods (FSS). Using a ML model trained on DFT-PBE, they found a clear transition line and estimated the critical point of that model. Recently, another ML model trained on DFT-PBE data~\cite{tenti_hydrogen_2025}, found supercritical behavior, i.e. a continuous dissociation with pressure with the location of the Widom line agreeing with the established DFT-PBE LLPT line.

The goal of the present study is to provide a more realistic estimate of the phase diagram of hydrogen in the region of the metal-insulator and molecular-atomic transitions. The previous studies have several major problems. The first requirement of such a calculation is to have an accurate proton-proton potential. Although a DFT functional can be accurate, the  PBE functional is likely not \cite{Clay2014}. Both the LLPT line and the melting line can vary considerably when changing functional and with respect to QMC predictions~\cite{niu2023stable}.
Secondly, a critical point can only be identified by accurate simulations comparing systems with different number of atoms so that one can see that the heat capacity and/or the compressibility diverge in the thermodynamic limit below the critical point but not above.  Although such studies could be accomplished directly with QMC or another accurate electronic structure method, direct calculations for systems with thousands of protons in the region of pressure and temperature where critical fluctuations are present are very costly. Using ML potentials trained on accurate QMC energies and forces allow such finite-sized studies to be accomplished.  Finally, the protons at these conditions have significant quantum motion, so estimates of its critical properties require path integral simulations in the pressure-temperature range of the LLPT.

\begin{figure}[h]
  \centering
  \includegraphics[width=\linewidth]{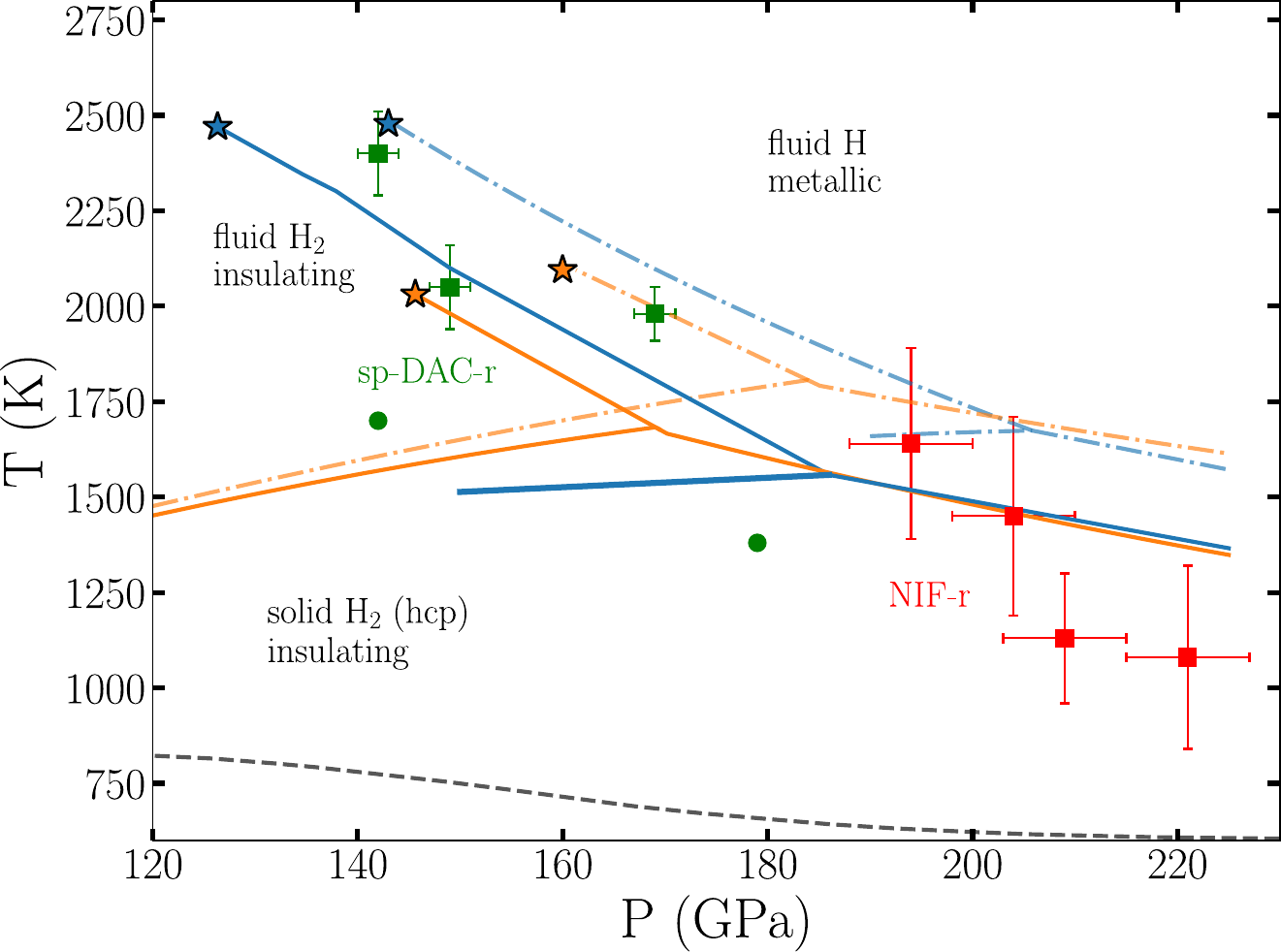}
  \caption{Phase diagram of hydrogen around the LLPT.
  Solid lines are predictions for quantum protons while the dot-dashed lines are for classical protons. \DMC{Orange lines indicate the LLPT and melting transition from a molecular hcp lattice (Phase-I) of the M29 model while blue lines indicate the same for the  D2508 model.}
  The estimated critical points are shown as stars. 
   Points show results of shock experiments at NIF (red squares, deuterium) ~\cite{celliers2018insulator}, DAC measurements (green circles, hydrogen)~\cite{Dzyabura_PNAS_2013} and (green squares, deuterium)~\cite{zaghoo2018striking}.
  These points indicate the threshold of the reflectivity $R=0.3$ associated with the insulator-metallic transition and the LLPT. \DMC{The black dashed line is the experimental melting line of hydrogen~\cite{Zha2017} inferred from changes in the measured Raman spectra.}
   } 
  \label{fig:Phase3}
\end{figure}

In this work we perform molecular dynamics (MD) and path integral MD calculations (PIMD) using two different machine-learned (ML) force fields trained with QMC data to explore the phase diagram of liquid hydrogen in the region of 100-200~GPa and 1500-3000~K. This allows us to estimate the influence of the ML potential on the phase diagram and other properties. Both models predict a first-order phase transition from the molecular to the atomic liquid terminating in critical points. 
Our best estimate of the phase diagram with comparison to experimental reflectivity measurements as a fingerprint of molecular dissociation and metallization, is shown in Fig.~\ref{fig:Phase3}. 

\section{Methods \label{sec:LLPT}}
In this section we outline the methods we used to estimate the liquid-liquid transition in dense hydrogen, in particular  how we locate the critical point using finite-size scaling and histogram reweighting. 
\subsection{Configuration selection}

The machine learning models were trained on 13,800 configurations of 96 protons in periodic boundary conditions, covering pressures from 50 GPa to 200 GPa and temperatures from 1600 K to 4000 K.
In order to guard against a bias for or against one of the phases 
the number of training and testing configurations  were split evenly between the three phases: solid molecular hydrogen, liquid molecular hydrogen and liquid atomic hydrogen. 

\subsection{Quantum Monte Carlo Methods}
We used Diffusion (DMC) and Variational (VMC) Quantum Monte Carlo methods to determine the electronic energy and forces of the configurations that went into training our machine learning models. We computed the electronic ground state energies and forces using the QMCPACK~\cite{Kim2018,Kent2020} software. We assumed the trial wavefunction was of the Jastrow-Slater form which is accurate for many-body hydrogen in this range of density~\cite{pierleoni2008}. The Slater determinant was taken from a DFT calculation with the PBE functional. The 2-body Jastrow factor was optimized for each configuration.
We assume the Born-Oppenheimer approximation for all our configurations. Electrons are in their ground state for temperatures and densities relevant to the LL critical point since the Fermi energy of the atomic liquid for densities near the critical density (0.45~\AA$^{-3}$) is roughly  $2 \times 10^5$K. We can safely say that electronic excitations are negligible for the thermodynamic properties.  
We used the Chiesa et al. estimator~\cite{Chiesa2005} to compute the forces on the protons. 

\subsection{Machine-Learning models}

We used message-passing graph neural network interatomic potentials MACE~\cite{batatia2022mace} and DeePMD~\cite{zhang2026graphneuralnetworkera} to run large scale MD simulations. 
In these models, configurations are mapped onto a set of learnable features, which in turn are mapped to the total energy. The features associated with a given atom depend not only on the coordinates of its neighbors, but also on their neighbor's features. Feature construction proceeds recursively so that information at a larger distance is used. 

The two models were trained on the same QMC dataset.\footnote{The methodology of using QMC data to train ML models is discussed in ref.~\cite{ceperley2023training}.}
 We will refer to our particular MACE model as M29 and the DeePMD model as \dpm. Both are state-of-the-art machine learning models: MACE is a good representative of an equivariant machine learning model and DeePMD is a state-of-the-art representative of an invariant machine learning model~\cite{peng2026lambench, kovacs_evaluation_2023}. 
MACE differentiates itself from other equivariant message-passing architectures by explicitly incorporating higher-body interactions within the message-passing process. 
The DeePMD model uses invariant features and incorporates higher-body effects by applying line-graph transformation to create higher order graphs whose vertices correspond to higher-body terms and then does message passing over that hierarchy~\cite{zhang2025graph}. 
Hyperparameters such as interaction cutoffs were tuned by minimizing energy and force errors of a test set. Since DMC energies are much more accurate than forces and dictate thermodynamic properties, we prioritized the energy root-mean-square-error (RMSE) with respect to the force RMSE.

Evaluated on the test set, the M29 model has a RMSE of energy 4.8 meV/atom and a RMSE of each force component of 129.9 meV/\AA~; the \dpm~model has a RMSE of 5.3 meV/atom and 131.5 meV/\AA~respectively.

\subsection{Determination of phase behavior}
\label{sec:phaseb}
We investigated the phase diagram using molecular dynamics (MD) in the NPT ensemble. We employed the LAMMPS~\cite{LAMMPS} software when considering classical protons and used LAMMPS built-in PIMD~\cite{li2026fix}  and the i-PI software~\cite{iPI3} for simulating quantum protons.
Nuclear exchanges were neglected as they are irrelevant for the conditions under investigation.

\begin{figure}
    \centering
    \includegraphics[width=\linewidth]{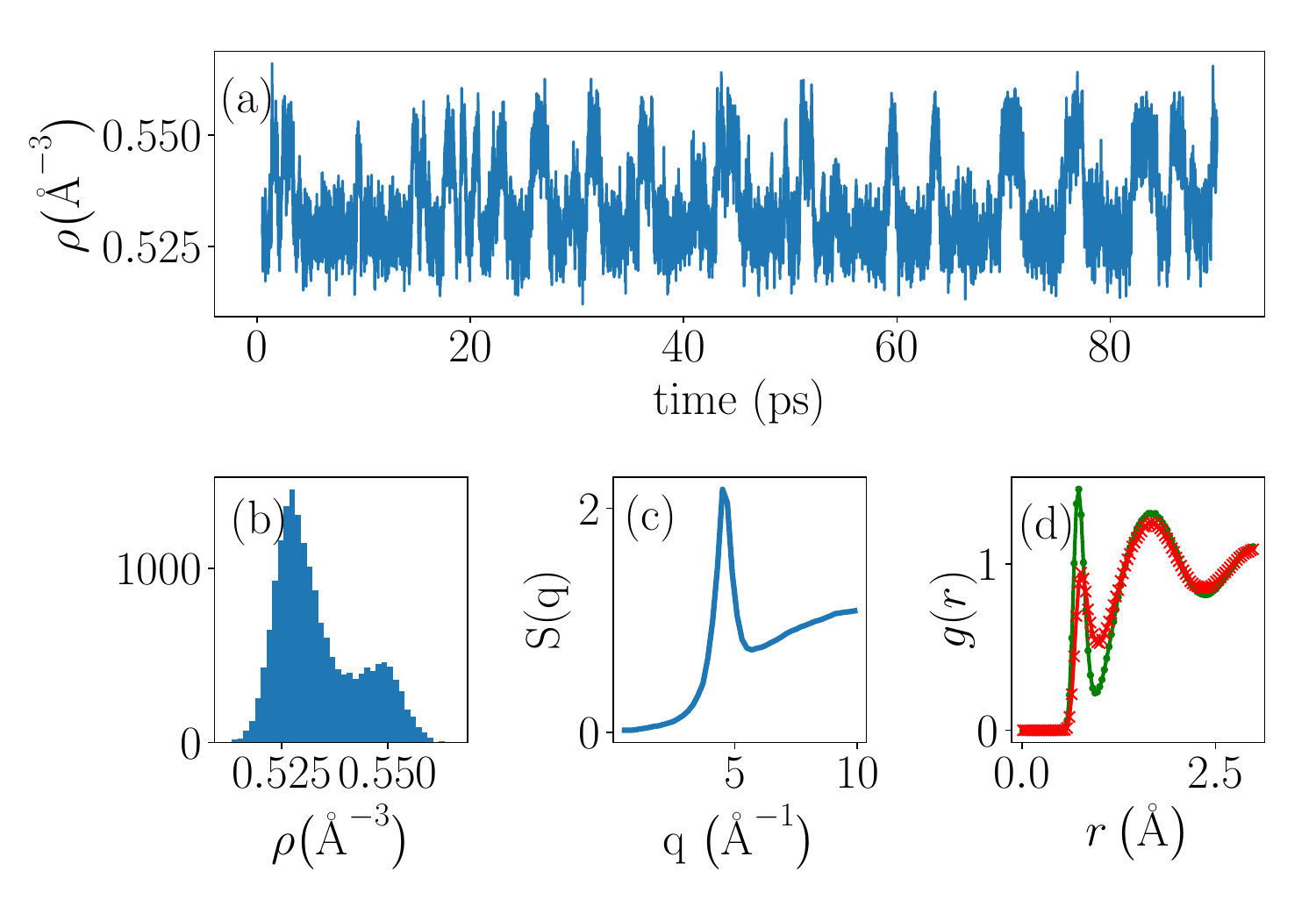}
    \caption{(a) Instantaneous density of a MD trajectory as a function of time with the M29 potential for 576 atoms at $193$ GPa  and $1700K$. (b) The histogram of densities during the trajectory. (c) The structure factor averaged on the full trajectory does not indicate any broken translational order, consistent with a liquid phase. (d) The average radial distribution function for low  ($\rho < 0.54~\mathrm{\AA}^{-3}$, green dots) and high ($\rho > 0.54~\mathrm{\AA}^{-3}$, red crosses) density configurations.
    }
    \label{fig:traj_npt}
\end{figure}

Due to the remarkably high temperature melting lines predicted in recent studies~\cite{niu2023stable,goswami2025high}, \DMC{at variance with earlier theoretical and experimental indications \cite{Bonev2004,morales2010evidence,Zha2017,Luo2026}}, it is necessary to check that the observed transition is liquid-liquid and not solid-liquid. 
The absence of Bragg peaks in the structure factor, as illustrated in Fig.~\ref{fig:traj_npt}(c), shows the liquid character of phases on both sides of the transition.
See Table (\ref{tab:melting}) for upper and lower bounds of the melting temperatures of the two models as a function of pressure.

To establish the first-order nature of the LLPT transition, we looked at the discontinuity in the equation of state (EOS), and the divergence of the compressibility and the heat capacity, with increasing system size. 
Other properties accompanying this transition and further details of our NPT MD runs are provided in the supporting information (SI).

The first indication of a first-order transition is the observed bimodality in density fluctuations at constant pressure in the transition region. In Fig.~\ref{fig:traj_npt}(a) we show the instantaneous density of a MD trajectory for a system with 576 classical protons as a function of time at 193 GPa  and T=1700K for the M29 model. The density fluctuates between the molecular and atomic phases roughly every few ps. The minimal effect of hysteresis occurs at all temperatures, for both classical and quantum MD. The density distribution is bi-modal  as shown in Fig.~\ref{fig:traj_npt}(b). 

This character change, from predominantly molecular to predominantly atomic, is indicated by the reduction of the molecular peak in the radial distribution function, as seen in Fig.~\ref{fig:traj_npt}(d). The radial distribution function in the low-density molecular phase exhibits a sharp peak at roughly 0.7~{\AA}, the characteristic bond-length of H$_2$ molecules. This peak is significantly reduced for the higher-density atomic phase, indicating a much reduced molecular character. 

\begin{table}[h]
    \centering
        \caption{Estimates of the melting temperatures (in K) of classical and quantum hydrogen with the M29 and \dpm\  models vs. pressure (GPa). The upper (u) and lower (l) bounds are determined for classical hydrogen using the 2 phase method with 2048 atoms. The quantum estimates are from using a method similar to that in ref. \cite{goswami2025high}. Details are given in the SI. The bounds for the M29 model are consistent with those of the M18 model from ref.~\cite{goswami2025high}. \DMC{The \dpm\ model melts at about 100K lower temperature than does the M29 from the molecular liquid but only 15K lower from the atomic liquid.} }
    \begin{tabular}{|c|c|c|c|c|c|}
    \hline 
        P  & $T_{l}^{M29}$  & $T_{u}^{M29}$  & $T_{l}^{\dpm}$  & $T_{u}^{\dpm}$ &$T_Q^{M29}$\\
        \hline
        150  &1645 &1660 &1560 & 1570&1603\\
        175  &1780 &1790 &1640& 1650& 1634\\
        200  &1700&1720&1690&1700&1478\\
        225  &1620 & 1625&1590&1620&1358\\
         \hline
    \end{tabular}
    \label{tab:melting}
\end{table}

\begin{figure}
    \centering
    \includegraphics[width=\linewidth]{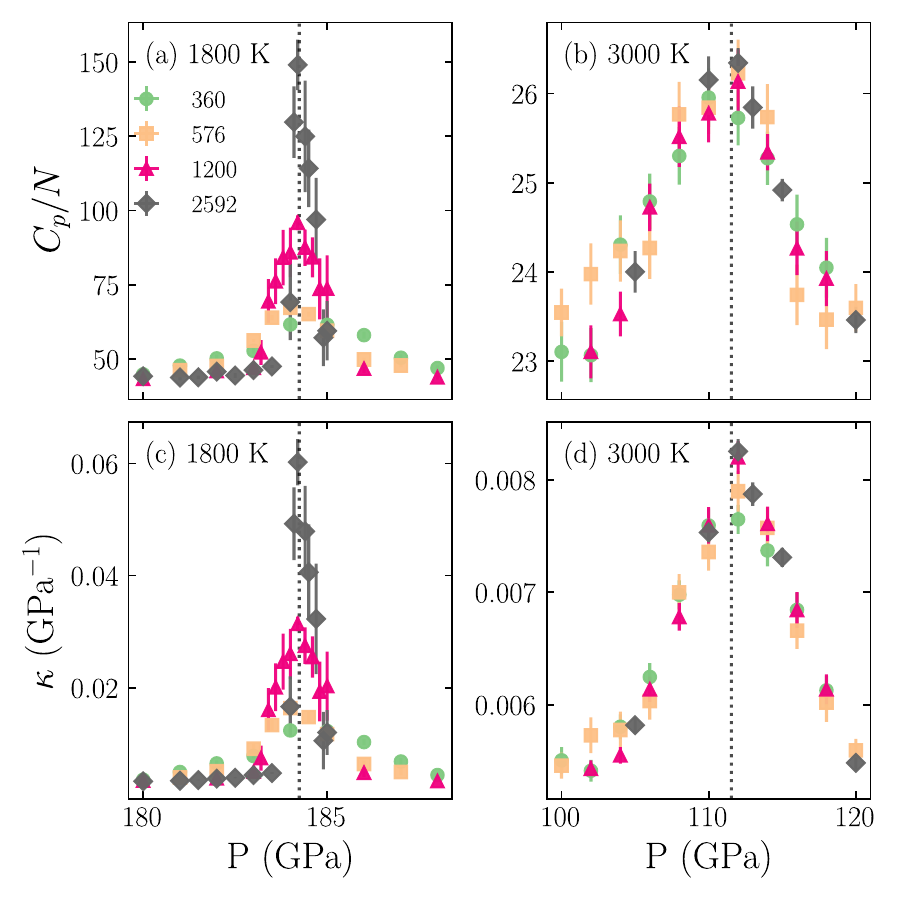}
    \caption{ Heat capacity at constant pressure $C_p$ and isothermal compressibility $\kappa$ for the M29 model. \DMC{The separate symbols designate the number of atoms.}
     The vertical dashed lines show the transition pressure obtained from Eq. (\ref{eq:EOS-tanh}). Panel (a) shows a phase transition at 1800 K as $C_p$ gets sharper and higher with system size. Panel (b) shows a crossover at 3000~K as $C_p$ is independent of system size. Panels (c) and (d) show that the compressibility exhibits the same behavior as in panels (a) and (b) for $C_p$.}
    \label{fig:Cp_4panel}
\end{figure}

The bi-modal behavior is not enough to differentiate a first-order transition from a crossover behavior. For this, we investigate the heat capacity, $C_p$, (computed from the enthalpy fluctuations) and the isothermal compressibility, $\kappa$, (computed from the volume fluctuations) with increasing system size as shown in Fig.~\ref{fig:Cp_4panel}. 
In the transition region, these properties exhibit a strong size dependence for temperatures below the critical temperature $T_c$ (Fig~\ref{fig:Cp_4panel}(a,c)), while they are independent of the system size for $T \geq T_c$ (Fig~\ref{fig:Cp_4panel}(b,d)). This is a clear signature of a first-order character of the transition below $T_c$, terminating in a critical point at $T_c$.

To estimate the location of the transition line we first fit the EOS (density versus pressure) along several isotherms and different system sizes to the phenomenological form~\cite{istas2025liquid, binder_finite_size_1984}:
\begin{equation} \label{eq:EOS-tanh}
    \rho (P) = \frac{D}{2}\tanh \left[ \alpha(P-P_t)\right] + \kappa_b (P-P_t) + \rho_t.
\end{equation}
Some of the fit parameters are reported in Table (\ref{tab:EOStable}). Note that the estimated relative density change across the LLPT at fixed temperature ($D/\rho_t$) is quite small: it is at most 4.0\% for M29 and 4.7\% for \dpm. 
\begin{table}[h]
    \centering
        \caption{Temperature, pressure, density and density jump along the LLPT. The average density from simulations of classical protons (Clas) and quantum hydrogen (H) at each temperature was fit using Eq. (\ref{eq:EOS-tanh}). Shown are $P_t$, $\rho_t$ and $D$  from those fits. 
    \label{tab:EOStable}}
    \begin{tabular}{|c|c|c|c|c|}
    \hline
model &T (K) & $P_t$ (GPa) &$ \rho_t$ (\AA$^{-3}$)&D (\AA$^{-3}$)\\
\hline
\DMC{M29} \DMC{Clas} & \DMC{1700} & \DMC{193.5} &\DMC{0.5391} &\DMC{0.0232} \\
M29 Clas &1800 &184.3& 0.5271& 0.0210\\
M29 Clas &2000 &168.4& 0.5053&0.0208\\
M29 Clas &2100 &161.1& 0.4947&0.0183\\
\hline
\DMC{M29} \DMC{H} & \DMC{1600} & \DMC{174.5} &\DMC{0.5168} &\DMC{0.0237} \\
M29 H &1700 &167.9& 0.5080&0.0242\\   
M29 H &1800 &161.1& 0.4973&0.0228\\ 
M29 H &2000 &149.1& 0.4798&0.0238\\ 

\hline
\dpm\ Clas &1800 &194.1 & 0.5420&0.0259\\
\dpm\ Clas &2100 &169.1 & 0.5075&0.0248\\
\dpm\ Clas &2400 &148.1 & 0.4749&0.0183\\
\hline
\dpm\ H &1800 &170.1 & 0.5115&0.0199\\
\dpm\ H &2100 &149.5 & 0.4807&0.0244\\
\dpm\ H &2400 &132.9 & 0.4537&0.0210\\
\hline
    \end{tabular}

\end{table}

\begin{figure}
    \centering
    \includegraphics[width=0.9\linewidth]{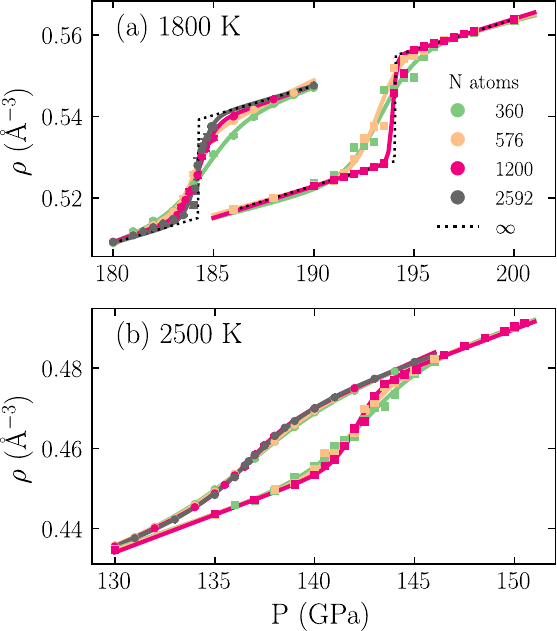}
    \caption{Density versus pressure and number of classical protons for the two models, \dpm ~(squares, higher pressure) and M29 (dots, lower pressure), at two different temperatures. (a) Both models show first-order phase transition behavior at 1800 K. (b)  M29 shows a cross-over at 2500K, whereas it is nearly critical for \dpm.}
    \label{fig:EOS_2panel}
\end{figure}

The behavior of the compressibility $\kappa$ corroborates the first-order nature of the transition. As shown in the EOS (Fig.~\ref{fig:EOS_2panel}), below the critical temperature (panel a), the compressibility diverges in the thermodynamic limit while above the critical point (panel b) it does not depend significantly on the system size.

Along each isotherm $T_0$, we obtain the density, $\rho(P)$, as a continuous function of pressure using histogram reweighting~\cite{binder_book_monte_carlo, ferrenberg1989optimized}. This allows us to calculate the maximum compressibility
\begin{equation}
   \kappa(N) = \max\limits_{P} \left( \frac{\partial\rho}{\partial P} \right)_{T=T_0}.
\end{equation}
To isolate the singular contribution at the transition line, $\kappa_t(N)$, we subtract a size and pressure independent background term, $\kappa_b$, calculated at pressures $8$ to $15$ GPa above critical point $P_c$ using N=1200 and 2592 atoms.

As shown in the upper panel of Fig.~\ref{fig:FSS_histogram_reweighting} 
the compressibility diverges linearly with respect to system size
at 1800 K for M29 and at 2200 K for \dpm. At temperatures  approaching the critical point this agreement with linear extrapolation deteriorates, indicating the breakdown of the first-order nature of the transition for both models.

\subsection{The location of the critical point\label{sec:critical_point}}

Although these observables show a clear sign of a first-order phase transition, the location of the critical point requires more refined finite-size scaling analysis described below. Table \ref{tab:PCTC} summarizes the predictions of the critical point based on different observables and extrapolation methods for both models. 

While the transition lines of our ML models are close to each other, the critical point location exhibits model dependence:$160$~GPa at $2100$~K for M29, in contrast to $140$GPa at $2500$K for \dpm.
Incorporating nuclear quantum effects further shifts the hydrogen transition line lower by approximately $30$~GPa for both models. 

We narrow the uncertainty in the location of the critical point by mapping the finite-size scaling onto the 3D universality class to obtain the correct scaling at the critical point.
Universality classes are determined by a small number of features: the system dimensionality, the range of the interparticle forces, and the number of components in the order parameter. Since our ML models have short-range interactions, our system is in three spatial dimensions and the order parameter, the density, is a scalar quantity, we assume that LLPT critical point for fluid hydrogen is in the 3D Ising model universality class. 

To map our transition to Ising, we replace the Ising order parameter, the average magnetization per spin, by the density fluctuation, i.e. $s \to \Delta \rho \equiv (\rho - \rho_c) $, where $\rho_c$ is the average critical density.
Within the finite-size scaling assumption~\cite{binder_finite_nodate}, the behavior of the distribution function $p_L(\rho, P)$ of a system of size $L$ in the immediate vicinity of the critical point is 
\begin{equation}
p_L(\Delta \rho, \Delta P) 
\sim L^{y} p(\Delta \rho L^{y}, L/\xi) e^{-\beta V \Delta P},
\label{eq:Fss_ansatz}
\end{equation}
where 
$\xi$ is the correlation length of the density fluctuations, $\Delta P=P-P_c$, and $p(\cdot,\cdot)$ is a universal scaling function.
In the vicinity of the critical point, the average density then behaves as
\begin{equation}
  \langle \Delta \rho \rangle = L^{-y} f(\beta \Delta P L^{3 - y} \rho_c^{y/3}, L/\xi),
  \label{eq:avg_delta_rho}
\end{equation}
where $f(\cdot,\cdot)$ is a function resulting from the integration 
over the above distribution function.
At the critical point, $\Delta P=0$, $L/\xi \to 0$. Using the Ising exponents with $y=\beta / \nu$, $\beta\simeq1/3$, $\nu \simeq 2/3$~\cite{binder_finite_nodate} and 
$\kappa_t(N) \sim \frac{\partial \langle \Delta \rho \rangle}  {\partial P}$,
we find the scaling behavior
$\frac{\kappa_t(N)}{\kappa_t(N')} = \left(\frac{N}{N'}\right)^{2/3}$,
distinct from the linear dependence on $N$ characteristic of a first-order transition below $T_c$.

\begin{figure}[htbp]
  \centering
  \qquad
  \includegraphics[width=0.8\linewidth]{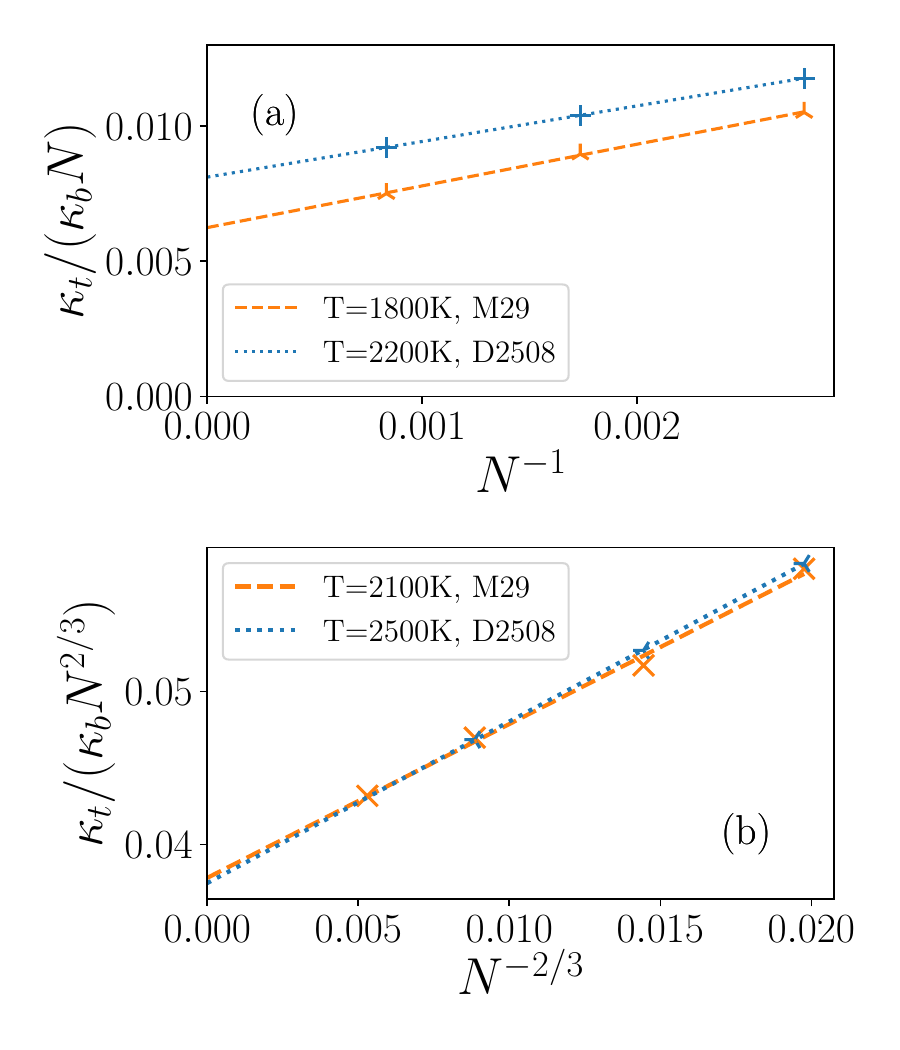}
  \caption{(a) Estimation of the dependence of the maximum compressibility $\kappa_t$ for M29 (blue) and \dpm (orange) versus system size below the critical temperature and (b) at the critical temperature. Points are estimates of the maximum compressibility from histogram reweighting for classical hydrogen.  
  To extrapolate to the thermodynamic limit, straight dotted and dashed lines are shown as guides to the eye. In the lower plot, the two lines overlap making them difficult to distinguish.
 }
  \label{fig:FSS_histogram_reweighting}
\end{figure}

Fig.~\ref{fig:FSS_histogram_reweighting} shows that at 2100K for M29 and 2500K for \dpm, both models follow the scaling 
$\kappa_t(N) \sim N^{2/3}$
as expected in the vicinity of the critical point from the finite-size scaling analysis.
This is especially visible in the M29 data for which we run one extra system size with $N=2592$ atoms.
Similarly, above 2300K for M29 and 2800K for \dpm, the extrapolated values vanishes, so that the change from molecular to atomic liquid becomes continuous. 
\begin{table}[h]

\begin{center}
\caption{Estimates of the critical point and triple point temperatures, pressures and densities for the two models from different methods. 
}
\begin{tabular}{ |c|c|c| }
\hline
  & M29 & \dpm \\
\hline
Critical point & & \\
\hline
 \makecell{Ising fit \\(quantum)} &
  \makecell{$2031 \pm 20~{\rm K}$ \\ $145 \pm 1.2~\rm{GPa}$ \\ $0.480 ~\angstrom ^{-3}$} &
  \makecell{$2470 \pm 9~{\rm K}$ \\ $127 \pm 0.5~\rm{GPa}$ \\ $0.467~\angstrom ^{-3}$} \\ 
  \cline{1-3}
                                 \makecell{Ising fit \\(classical)} &
  \makecell{$2095 \pm 9~{\rm K}$ \\ $160 \pm 0.7~\rm{GPa}$ \\ $0.483 ~\angstrom ^{-3}$ } &
  \makecell{$2478 \pm 7~{\rm K}$ \\ $143 \pm 0.5~\rm{GPa}$ \\ $0.461 ~\angstrom^{-3}$} \\
  \cline{1-3}
                                \makecell{$\kappa_t(N)$ scaling \\ (classical)} & 
   \makecell{$2100 \pm 200~{\rm K}$ \\ $160 \pm 10~\rm{GPa}$ \\ $0.484 ~\angstrom^{-3}$} &
   \makecell{$2500 \pm 200~{\rm K}$\\ $140 \pm 10~\rm{GPa}$ \\ $0.453 ~\angstrom^{-3}$ }\\ 
\hline
\hline
Triple point & & \\
\hline
 hydrogen & 
    \makecell{\DMC{$1678$ K}\\ \DMC{$169.4~\rm{GPa}$} 
    }&
    \makecell{$1558$ K\\ $186~\rm{GPa}$ \\ 
    }\\
    \hline
 deuterium & 
    \makecell{\DMC{$1735$ K}\\ \DMC{$177.7~\rm{GPa}$} 
    }&
    \makecell{ \\  \\ 
    }\\
\cline{1-3}
                             classical & 
    \makecell{\DMC{$1805$ K}\\\DMC{$184.0~\rm{GPa}$}  \\
    }&
    \makecell{$1673$ K\\$204~\rm{GPa}$  \\ 
    }\\
    \hline
\end{tabular}
\label{tab:PCTC}
\end{center}
\end{table}

To estimate the critical temperature and pressure more precisely without measuring very small density differences, we use the field mixing technique developed by Wilding et al.~\cite{wilding_simulation_1997}.
It is considered to be the most precise method to find the critical point of liquid-gas transition~\cite{binder_book_monte_carlo}: it has been applied to the Lennard-Jones liquid-vapor phase transition~\cite{wilding_finite-size_1996}, and to the liquid-liquid phase transition in water~\cite{debenedetti_second_2020}.

We compare the probability distribution of the order parameter for fluid hydrogen with that of the Ising model; in the critical region they converge to the same universal function~\cite{tsypin_probability_2000}.
The distribution of the magnetization of the 3D Ising model at the critical point can be approximated as~\cite{tsypin_probability_2000}
\begin{equation}
    p_\text{Ising}(M) \propto \exp\left(- (M^2 / M_0^2 - 1 )^2 (a M^2 /M_0^2 + c) \right),
    \label{eq:Ising_probability}
\end{equation}
with $a = 0.158$, $c=0.776$, and $M_0 \propto L^{-\beta/\nu}$ with
$\beta / \nu = 0.518$~\cite{ferrenberg1991critical, binder_finite_nodate}.

Wilding et al.~\cite{wilding_simulation_1997} showed that to obtain an accurate comparison of the fluid critical fluctuations with the Ising model, identifying the magnetization with the fluid density suffers from large next to leading order effects.
In a system that lacks symmetry between phases, an improved mapping of the order parameter is given by $\delta m = m - \langle m \rangle$, where $m = \rho + s e$, $e$ being the potential energy per particle, $\rho$ the particle density and $s$ a scalar quantity controlling the degree of field mixing.

Using multiple histogram reweighting~\cite{binder_book_monte_carlo, ferrenberg1989optimized, newman1999monte}, we determine the corresponding probability distribution and the field mixing parameter $s$ of our hydrogen fluid containing $N$ protons,
$p_H(\delta m | P, T, N)$ for $(P, T)$ values close to the suspected critical point.
Precisely at the critical point, $p_H(\delta m|P_c,T_c,N)$
will approach the universal distribution, $p_\text{Ising}$, in the limit of large systems.
We therefore determine the critical pressure, $P_c(L)$, temperature, $T_c(L)$, for each system size $N \sim L^3$
by maximizing the overlap between~\eqref{eq:Ising_probability} and $p_H(\delta m | P, T,N)$.

\begin{figure}[htbp]
    \centering
    \includegraphics[width=0.9\linewidth]{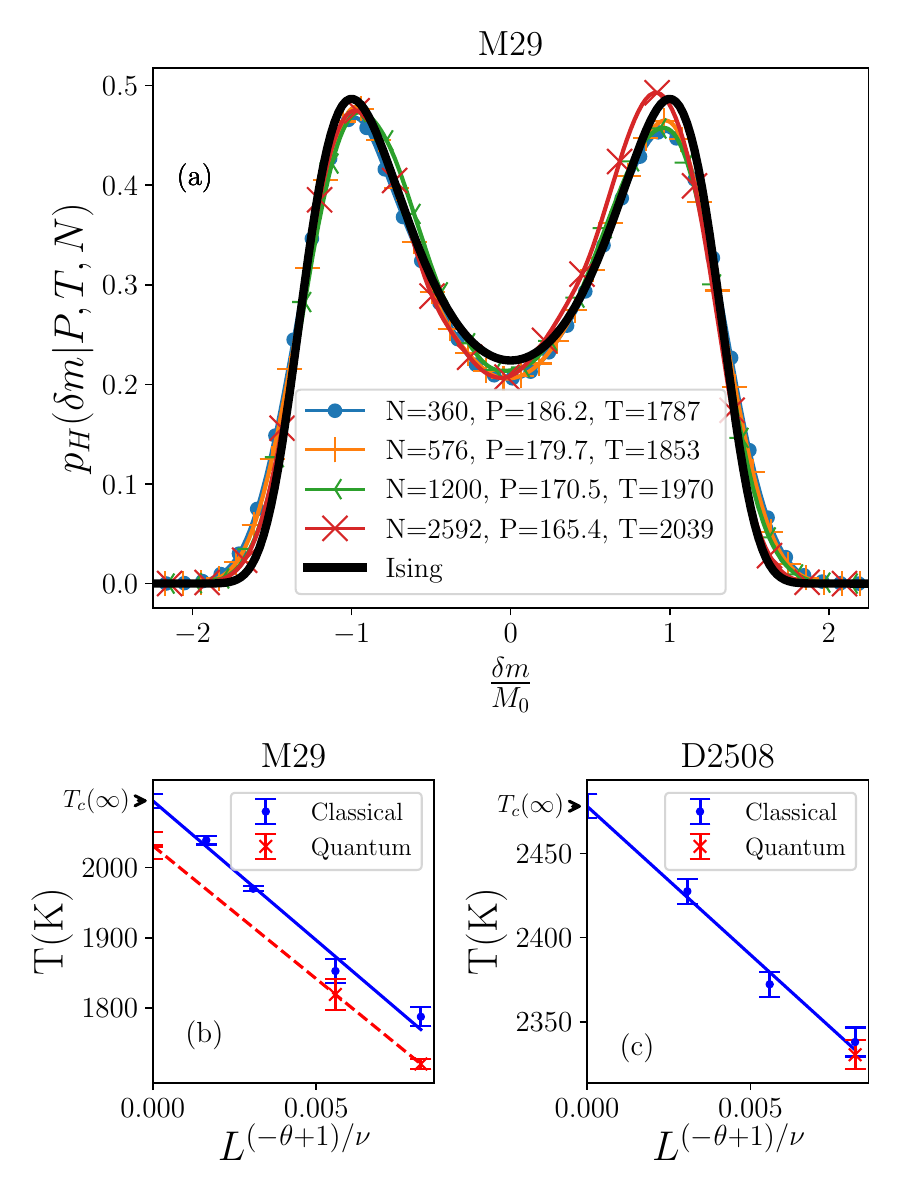}
    \caption{Procedures to estimate the LLPT critical point of hydrogen. (a) The order parameter distribution for the M29 model of classical hydrogen compared with the magnetization distribution of the 3D Ising model at criticality, given by Eq.~ (\ref{eq:Ising_probability}). The field mixing value is fixed to $s=-0.35$ ${\rm eV}^{-1} \angstrom^{-3}$ and is not sensitive to the system size; the proportionality constant $M_0 \propto L^{-\beta/\nu}$ is fixed for one system size to match with the Ising distribution, 
    $\beta / \nu = 0.518$~\cite{ferrenberg1991critical}.
    (bottom) Critical temperatures $T_c(L)$ for the M29 model (b) and the D2508~ model(c) as a function of system size. Critical pressures and temperatures are extrapolated to the infinite system based on the critical exponents $\theta = 0.54 \pm0.05$\cite{chen1982unbiased, wilding1995critical} and $\nu=0.629$~\cite{chen1982unbiased}. The blue points and continuous lines are for classical hydrogen, the red crosses and dashed lines for quantum hydrogen. 
    }
    \label{fig:M29_order_parameter_fit}
\end{figure}

Results are shown in fig.~\ref{fig:M29_order_parameter_fit}.
The residual dependencies of the critical temperatures, $T_c(L)$, on the system sizes are shown in the lower panels of Fig.~\ref{fig:M29_order_parameter_fit}.
The apparent reduced critical temperature~\cite{wilding_simulation_1997} converges as $|T_c(L) - T_c(\infty)| \propto L^{-(\theta + 1) / \nu}$ leading to $T_c(\infty)\simeq 2478 K$ for the \dpm ~ model and to $T_c(\infty)\simeq 2095 K$ for the M29 model.
The estimated critical pressures $P_c(\infty)$ are then determined from the LLPT line at  $T_c(\infty)$.

\subsection{Quantum proton effects}\label{sec:quantum_effects}
We used PIMD to calculate the density fluctuations for quantum protons. We estimated the quantum critical point using a generalization of the histogram reweighting and finite-size  scaling analysis described above: nuclear quantum effects do not modify the universality class at finite temperature. The changes in the critical point are shown in the lower panels of Fig.~\ref{fig:M29_order_parameter_fit}. We find that quantum effects move the transition line of hydrogen to lower pressures at a given temperature. The critical point is approximately ($145$ GPa, $2031$ K) for the M29 model and ($127$ GPa, $2470$ K) for the \dpm\  model as shown in Fig.~\ref{fig:Phase3}.  Details of the procedures for histogram reweighting for path integrals are given in the SI. 

\subsection{Triple point}\label{sec:triple point} 
The intersections of the classical and quantum M18 melting lines~\cite{goswami2025high} with the corresponding LLPT lines are used to determine  the  triple point,
$(T_3, P_3)$ , where the molecular hcp solid (S), molecular liquid (ML), and atomic liquid (AL) phases coexist. Results are given in table~\ref{tab:PCTC}.
The slope of the solid $H_2$-liquid $H$ 
(S-AL) phase boundary, drawn in the phase diagram of fig.~\ref{fig:Phase3}, follows from the Clausius-Clapeyron (CC) relations at the triple point among the three phases by using the slopes of the S-ML melting line, its volume change, the slope of the ML-AL line and its volume change. 
We have checked the melting temperatures of the two models with 2-phase studies of classical hydrogen  with results consistent with the CC estimates: see Table \ref{tab:melting}. The location of the triple point is in agreement with the estimate in ref.~\cite{goswami2025high} for the  M18 model. \DMC{ Away from the critical point we estimated the quantum melting temperatures of hydrogen and deuterium monitoring the derivative of the melting temperature with respect to the ionic mass. Details are given in the SI.}

\section{Summary and Outlook}

We predict the phase diagram of dense hydrogen using two machine-learning models as shown in Fig. (\ref{fig:Phase3}).  We find a first-order phase transition in both models extending above the hcp molecular solid.  
But the two ML models differ in the precise location of the LLPT:
at a fixed temperature, the difference between the two models is about 5\% in the transition pressure and about 2.7\% in the transition density. The difference in their critical temperatures is $\approx 18\%$. 

The large difference in the critical point of the two ML models could arise from how the proton-proton interactions for distances larger than occur in the training data are treated in the ML models.  Although size consistency is guaranteed in both ML models, use of ML energies and forces for systems larger than those used for training introduces an uncertainty which needs to be verified. We are currently investigating this by using data for larger systems in the training set. We note that smaller differences, $\approx 6\%$, are found in the melting temperatures of the two models: see Table \ref{tab:melting}.

In the SI we compare the current results with results from previous simulations using QMC and FP-DFT simulations \cite{scandolo2003liquid,morales2010evidence,Bonev2004,Tamblyn2010,lorenzen2010first,karasiev2021liquid,istas2025liquid,tenti_hydrogen_2025,Morales2013,geng2019thermodynamic,Hinz2020,Bund2021,bergermann2024,Mazzola2018,tirelli2022high} 
The difference between the current ML models is much less than with those of DFT predictions based on different XC-functionals. For example  changing the DFT functional from PBE to vdW-DF1 changes the transitions/Widom lines by $\sim 70$~GPa. The melting line of PBE-DFT is a factor of 2 lower than that obtain with the QMC trained model.
We attribute the higher melting temperatures in the QMC trained models to the DFT models to the attractive dispersion forces of the molecular crystals that are not well approximated by the PBE-DFT functional. See ref.~\cite{goswami2025high} for a comparison. 

A consequence of the liquid-liquid first-order transition is the presence of a triple point on the melting curve.  The dividing line between a metal and an insulator (MI) starts at the critical point, reaches the triple point and then continues along the melting line. At higher pressures the MI line follows the molecular melting line until either the molecular crystal becomes unstable or until it becomes metallic.  
The triple point could be a clear experimental indication of the LLPT that would be detected in ramp compression studies. The melting temperature will jump from increasing to decreasing with respect to pressure at the triple point caused by the higher density of the atomic liquid with respect to the molecular liquid. Without a first-order transition, the change under equilibrium conditions would be more gradual.  

We expect that the LLPT for deuterium and its triple point will lie roughly halfway in between the hydrogen phase lines and the classical ones.
We have tested this expectation using PIMD simulations of deuterium. See the SI for details.

Experimental observations of the metallic state in a dense fluid have been reported by dynamical compression
experiments \cite{weir1996metallization,fortov2007phase,mochalov2017quasi,knudson2015direct,celliers2018insulator} and static compression methods heated by laser pulses
\cite{Dzyabura_PNAS_2013,zaghoo2016evidence,zaghoo2018striking,Jiang2020}, yielding controversial results.
 Fig. \ref{fig:Phase3}  shows comparison of the shock experiments of NIF \cite{celliers2018insulator} as well as dynamic DAC experiments \cite{Dzyabura_PNAS_2013,zaghoo2016evidence} that probed the pressure-temperature regime near the LLPT. We note that the NIF experiments do not directly measure temperature, the systems are dynamic on small inhomogeneous samples. 
 The LLPT pressures is assigned to when the reflectivity reaches 30\%. This corresponds to a minimum metallic conductivity of 2000 S/cm for bulk systems~\cite{Rillo2019} indicative of the MI transition.
 In ref.~\cite{zaghoo2016evidence} the temperature is inferred from a black-body radiation fit of the emitted light.
 Only integrated optical properties are observed.   They are suggestive of the MI transition but by themselves cannot establish whether the transition is continuous or weakly first-order.

In conclusion, the ML models based on accurate Quantum Monte Carlo calculations predict a  weak first-order transition between the insulating molecular liquid and the metallic atomic liquid. This transition is very near the melting transition. The sensitivity of the phase diagram to small changes in the proton-proton interaction leads to widely varying theoretical and experimental findings of the phase diagram.  A reliable prediction requires an accurate and realistic treatment of molecular and atomic interactions, protonic quantum effects, and a careful examination of finite-size effects in the critical region and of the melting temperature.

We plan detailed exploration of thermodynamic properties in this region of the phase diagram.
This will include experimental signatures such as Raman spectra \cite{Subramanian2011}  valuable for identifying the crystal structure. Direct comparison with experiment is of critical importance in validating the ML procedures and models.    We also plan on extending these calculations to pressures between the triple point and the solid molecular --solid atomic transition.   

\FloatBarrier
\section*{Acknowledgments}

Work by DC and SG was supported by the U.S. Department of Energy (DOE), Office of Science, Basic
Energy Sciences (BES) under Award DE-SC0020177. HN was supported by NSFC under grant number 12504278 and 12611541151. YY was supported by the National Science Foundation (NSF) under grant number DMR-2532734. CP and IS were supported by the European Union - NextGenerationEU under the Italian Ministry of University and Research (MUR) projects PRIN2022-2022NRBLPT CUP E53D23001790006 and PRIN2022-
P2022MC742PNRR, CUP E53D23018440001.

An award of computer time was provided by the INCITE program using resources of the Argonne Leadership Computing Facility,  a DOE Office of Science User Facility supported under Contract DE-AC02-06CH11357. 
We also used the GRICAD infrastructure 
supported by the Grenoble research communities, Illinois Campus Cluster \DMC{and the Delta advanced computing and data resource}, supported by the National Science Foundation (Awards No. OCI-0725070, No. ACI-1238993 \DMC{and award OAC 2005572}) the State of Illinois, the University of Illinois at Urbana-Champaign, and its National Center for Supercomputing Applications. \DMC{Delta is a joint effort of the University of Illinois Urbana-Champaign and its National Center for Supercomputing Applications. This work used Delta at National Center for Supercomputing Applications through allocation PHY240060 from the Advanced Cyberinfrastructure Coordination Ecosystem: Services \& Support (ACCESS) program}~\cite{NSF_ACCESS_reference}, \DMC{which is supported by U.S. National Science Foundation grants \#2138259, \#2138286, \#2138307, \#2137603, and \#2138296.}
We acknowledge EuroHPC JU for awarding the project ID EUHPC E05 077 Extreme Access to the DCGP partition at Leonardo@CINECA, Italy and CINECA for providing computer time through the ISCRA-B project HP10B27XJC.





\section*{Supplementary Information:\\
The phase diagram of fluid hydrogen near melting from machine-learned interatomic potentials trained on quantum Monte Carlo data}
\author{S. Goswami, H. Niu, I. Saitov, M. Istas, Y. Yang, S. Jensen, M. Holzmann, C. Pierleoni, D. Ceperley}


\subsection*{Configuration selection}
\begin{figure}[htpb]
    \centering
    \includegraphics[width=0.7\linewidth]{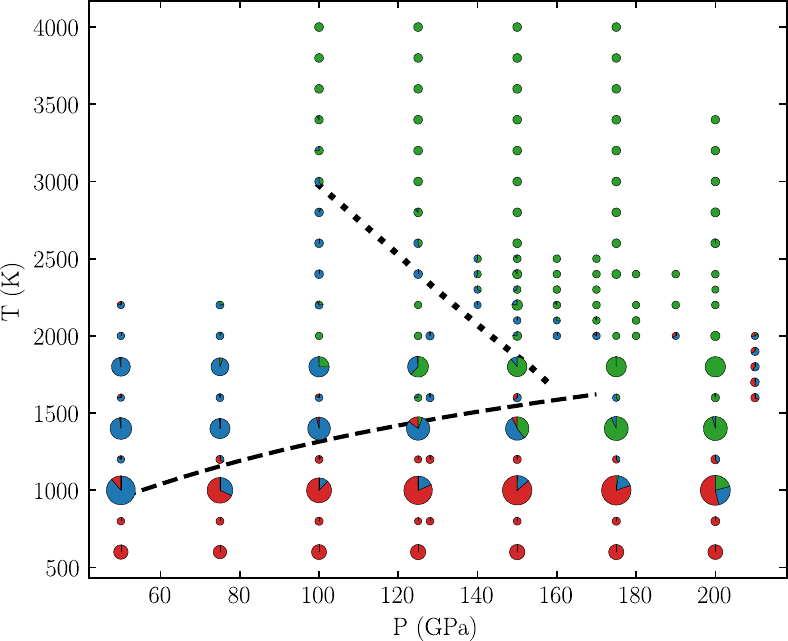}
    \caption{The pressure and temperature range of the training dataset of 13,800 configurations. The size of the circles correspond to the number of configurations at that phase point. The colors are an estimate of the phase of the configuration: atomic liquid (green), molecular liquid (blue)  or molecular solid (red). The molecular systems were identified by the bonding peak in the pair correlation function, the solids by a large peak in the structure factor.  The dashed line is the melting line estimate~\cite{goswami2025high} using M18 model. The dotted line is the estimated phase boundary between molecular liquid and atomic liquid in the M29 model.
    }
    \label{fig:LLPT_configs}
\end{figure}

The machine  learning models were trained on 13,800 configurations covering pressures from 50 GPa to 200 GPa and temperatures from 1600 K to 4000 K as shown in Fig. \ref{fig:LLPT_configs}. Note that hydrogen is a very compressible system: its density is quite sensitive to temperature and pressure. Our range of 50 GPa to 200 GPa already has the density increase 1.5 times. Since ML interatomic potentials have finite cut-offs in their interaction environments, wide density ranges for training data can bias the model toward higher-density phases that will contain more neighbors within the cutoff volume. Hence, we restricted our training data to have a density range relevant to what we will examine in this paper.
In addition to using some of the configurations from ref. \cite{niu2023stable} we generated new configurations using classical MD with the M18 MACE model ~\cite{goswami2025high}. All configuration had 96 protons. 

The number of training and testing configurations  were chosen to represent the three relevant phases equally: m-hcp solid molecular hydrogen (S), liquid molecular hydrogen (ML) and liquid atomic hydrogen (AL). This even splitting was designed to guard against a bias for or against one of the phases. 
The long-range strength of the interaction can change depending on whether the systems is a metal or insulator. 
The methods to characterize the phase of a given configuration was heuristic, based on their proton-proton pair correlation functions and structure factors. 
The precise choice and the potential used to sample configurations should not be important given that we have a sufficient coverage of configuration space in the training set and if generating model adequately represents the many-body potential. 

\subsection*{Quantum Monte Carlo Methods}

 For temperatures and densities relevant to the liquid-liquid phase transition, its critical point and the melting line, the electrons are in their ground state implying that the Born-Oppenheimer approximation is accurate.  The Fermi energy of the atomic liquid for densities near the critical density ($\approx 0.45$~\r{A} $^{-3}$) is roughly $2 \times 10^5$K, so we can safely assume electronic excitations are negligible for the thermodynamic properties. 

Diffusion (DMC) and Variational (VMC) Quantum Monte Carlo method were used to determine the electronic contribution to the energy and forces of the configurations in the training and test sets. We computed the ground state energies and forces using the QMCPACK code\cite{Kim2018,Kent2020}. We assumed the trial wavefunction was of the Jastrow-Slater form which is accurate for many-body hydrogen in this range of density\cite{pierleoni2008}. The Slater determinant was taken from a DFT calculation with the PBE functional. The 2-body Jastrow factor was optimized for each configuration.  We corrected for the electronic finite-size effects by applying twist averaging with $5^3$ twists values for the single electron contribution and by using the electronic structure factor for the many electron contribution to the energy\cite{holzmann2016theory}.  The typical statistical RMSE on the DMC total energy was 11 meV.
We used the Chiesa et al. estimator~\cite{Chiesa2005} to compute the forces on the protons. Extrapolation between the VMC and DMC forces was used to correct the mixed-estimator bias. We did not apply a finite-size correction to the forces.  The typical RMSE on each force component was 77~meV/{\r{A}}.

\subsection*{Molecular Dynamics Simulation Methods}
All classical molecular dynamics (MD) simulations of hydrogen were performed with LAMMPS package~\cite{LAMMPS}.  
Simulations employed three-dimensional periodic boundary conditions.
Interatomic forces were evaluated with the MACE-based M29 machine-learned potential and DeepMD's DPA3-based D2508 potential, implemented through the ML-IAP unified interface in LAMMPS. Neighbor lists in LAMMPS were built with a 2.0 A skin and the list capacities were increased relative to LAMMPS defaults to accommodate the high density neighborhood of hydrogen. 

We performed NPT MD simulations on four different system sizes namely 360, 576, 1200 and 2592 atoms. 
At NPT conditions, the volume of the simulation box is a dynamic quantity but we impose a orthorhombic box geometry throughout the simulation to prevent box collapse for the liquid states. The time step was set to 0.5 fs and the simulations ran up to 100 ps. To maintain thermodynamic conditions, we employed a Nosé-Hoover thermostat and barostat with damping constants of 50 fs and 500 fs, respectively. 
All classical simulations were executed on NVIDIA A40 GPUs using the Kokkos accelerator package. 

\subsubsection*{Two-phase simulations}
\begin{figure} [htpb]
\centering
\includegraphics[width=\linewidth]{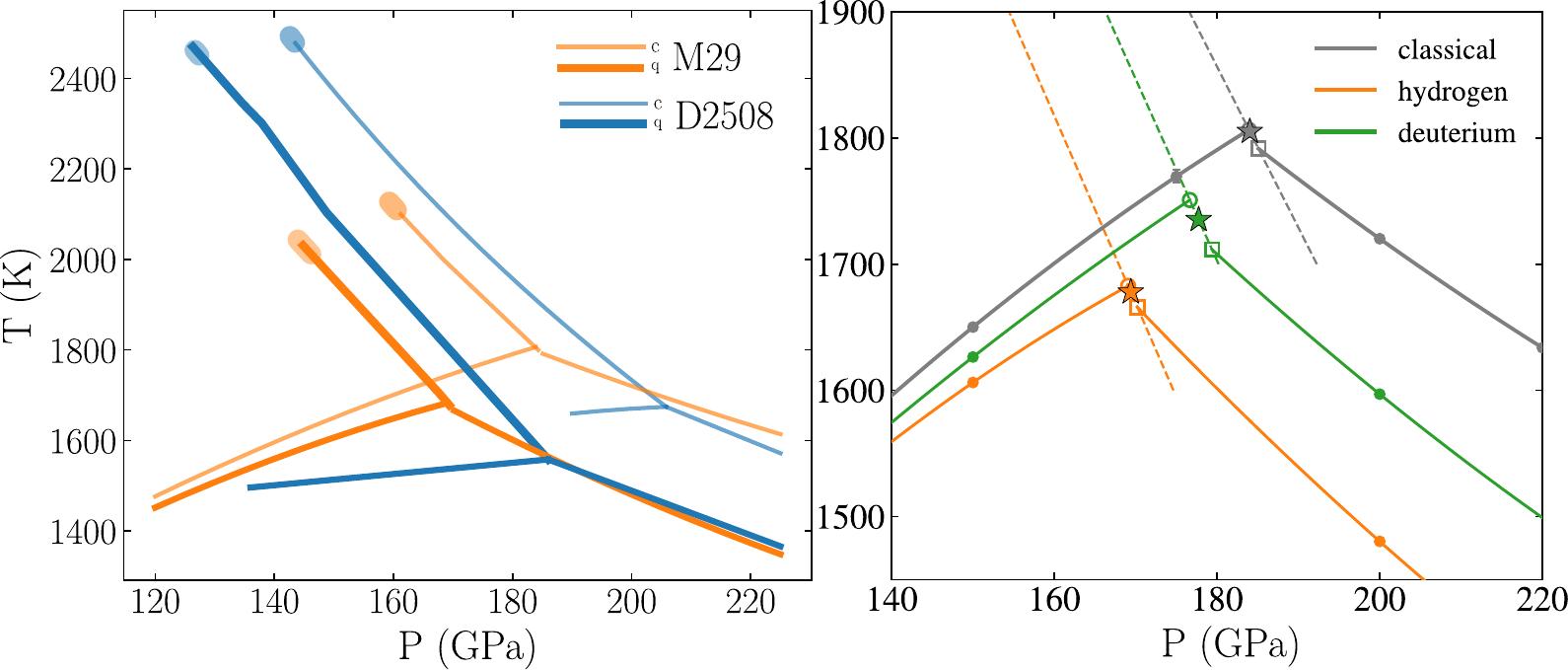}
\caption{\DMC{Left: The phase diagram (temperature vs. pressure) of classical (thin lines)  and quantum hydrogen (thick lines) estimated by the two ML models: M29 (orange) and \dpm (blue). Filled ovals show estimates of the critical points. Right: The M29 model triple point estimates (stars) for classical protons (gray), hydrogen (orange) and deuterium (green). Dashed lines are estimates of the LLPT. Solid lines are quadratic fits to the melting lines. Points on those lines are where estimates using Eq. (1) were performed. } }
\label{fig:LLPT-line}
\end{figure}

For two-phase simulations we employed 2048 atoms, half of which were initially liquid and half m-hcp solid. We created the configurations by first constructing two 128 atom cubic boxes of homogeneous phases of solid and liquid classical hydrogen separately. Those boxes were replicated to form two 2x2x2 cubes each of a single phase. For the atomic liquid and corresponding molecular solid the two boxes were equilibrated at 225 GPa and 1600 K while for the molecular liquid and molecular solid they were equilibrated at 150 GPa and 1500 K.
These equilibrated boxes were then stacked on top of each other, thus forming boxes containing 2048 atoms half liquid and half m-hcp solid with a small gap at the interface.  These systems were then simulated for up to 10 ps under NPT conditions. The structure factor was monitored to see if the Bragg peaks tended to a large value or a small value thus determining an upper or lower bound respectively to the melting temperature. For the bounds for the two models see figure \ref{fig:LLPT-line} and Table 1 in the main text.

\subsubsection*{Path Integral Molecular Dynamics (PIMD)}
Hydrogen, the lightest element, has significant nuclear quantum effects. The thermal De Broglie wavelength of the protons $\Lambda=\hbar(k_B T m_p)^{-1/2}$ is 0.18~{\r{A}} at T=1500~K.  The quantum effect of the protons is important to describe the vibrational motion of H$_2$ 
because for the temperatures under consideration, a H$_2$ molecule is likely to be in its vibrational ground state and thus quantitized.
We neglect exchanges of protons as insignificant at these temperatures.

Nuclear quantum effects were included using LAMMPS's built-in fix pimd/langevin \cite{li2026fix} and using the code i-PI\cite{iPI3}. LAMMPS runs the entire ring polymer within a single LAMMPS invocation, with one MPI rank per bead. All beads were initialized from the same classical configuration. The NPT dynamics equilibrated each replica independently. The path-integral Langevin dynamics used the obabo integrator with a $\mathrm{PILE_L}$ thermostat ($\lambda = 0.5$ ,$\tau= 10$ fs) and a BZP isotropic barostat with damping parameter of 20~fs. The integration timestep was 0.2 fs. We ran close to 50-70~ps of simulation using 10 beads on 3 different atoms sizes, namely 360, 576 and 1200 atoms to construct the quantum LLPT line. Simulations were launched across 4 NVIDIA A100 GPUs on Delta supercomputer. 

\subsubsection*{Estimation of the melting temperature for quantum protons}
\DMC{Rather than use the two phase melting for quantum hydrogen, we used a modification of the method discussed in ref. \cite{goswami2025high}.  This allows smaller homogeneous simulations, and a more precise estimate of the effect of protonic zero-point motion. In the neighborhood of the triple point, we need to consider how changing the ionic mass changes the melting temperature under constant pressure conditions, instead of the dependence of the melting pressure at constant temperature considered in ref. \cite{goswami2025high}.
We can determine how the melting temperature $T(P_m,x)$ at a pressure $P_m$ changes with the mass of the ion $m$, (written in terms of $x=m_p/m$) by differentiating the Gibbs free energy with respect to  $x$. For a point on the melting curve:
\begin{equation}
 \frac{dT(P_m,x)}{dx} =\frac{T\Delta k(T,P_m,x)}{x\Delta H(T,P_m,x)}
\label{eq:cons_P}
\end{equation}
where $\Delta H(T,P_m,x)=H_l(T,P_m,x)-H_s(T,P_m,x)$ is the enthalpy difference per atom between the liquid and solid phase at constant pressure
and $\Delta k$ the similar difference in the kinetic energy per atom. 
To use this expression, we perform PIMD simulations on homogeneous liquids and solids for values of $(T,P_m,x)$ around the suspected melting temperature to determine the differences in kinetic energies and enthalpy. 
Assuming the melting temperature, $T_m(P_m,x)$, is a smooth function of x, we expand $T_m(P_m,x)= T_c + T_1 x + T_2 x^2$ where $T_c(P_m)$ is the classical transition pressure (i.e. from table 1 in the main text).
 $T_1 $ can be estimated from the semiclassical expression \cite{Wigner1932}: 
\begin{equation}
 T_1=\lim_{x\rightarrow 0}  \frac{dT_m(P_m,x)}{dx} =-\frac{\hbar^2}{24 m_p k_B^2 T_m}\frac{\Delta \left< F_i^2 \right> }{\Delta H(T_m,P_m,0)}.
\end{equation}
Here $\Delta \left< F_i^2 \right>$ is the difference of the mean squared force on a classical proton between the classical liquid and solid at melting.
The coefficient $ T_2$ is determined by fitting to the right hand side of Eq.~\ref{eq:cons_P} to the expression $dT/dx=T_1+2T_2 x$.
Care must be taken near the triple point to ensure that the liquid phases are all either atomic or molecular, i. e. the constant pressure scans do not cross the LLPT.
Table \ref{tab:meltp}  shows our estimates of these quantities versus pressure for the M29 model.
Table \ref{tab:melting_quadratic_fit} shows the fit parameters for the classical, deuterium and hydrogen melting lines for pressures close to the triple points based on the M29 model.
Figure S2 (left) shows the phase diagrams for the two ML models. Figure S2(right) shows in greater detail the effect of quantum motion of the ions for the M19 model near the triple point. }

\begin{table}[h]
    \centering
        \caption{\DMC{Estimates of the melting temperatures (in K) of classical protons ($T_C$), deuterium ($T_D$) and  hydrogen ($T_H$) with the M29 model vs. pressure (GPa). $T_1$ and $T_2$ are the expansion coefficients of the melting temperature in powers of $x=m_p/m$. }}
    \begin{tabular}{|c|c|c|c|c|c|}
    \hline 
        P  & $T_C$  & $T_1$  & $T_2$  & $T_D$ &$T_H$\\
        \hline
        125  &1507.6 & -27.6 & 0.76 & 1494 & 1481 \\
        150  &1650.2 & -53.8 & 9.64 & 1626 & 1606\\
        200  &1720.2 & -252.4 & 12.3& 1597 & 1480\\
        225  &1613.8 & -280.0 & 16.4& 1478 & 1350\\
         \hline
    \end{tabular}
    \label{tab:meltp}
\end{table}

\begin{table}[h]
    \centering
    \caption{\DMC{Quadratic fit parameters of the melting line: $T=p_0+p_1P+p_2P^2$. T in K, P in GPa.}}
    \begin{tabular}{|c|c|c|c|c|c|c|}
    \hline
         mass & $p_0^{\text{mol}}$ & $p_1^{\text{mol}}$ & $p_2^{\text{mol}}$ & $p_0^{\text{atom}}$ & $p_1^{\text{atom}}$ & $p_2^{\text{atom}}$  \\
         \hline
         classical &445.99 &10.818 &-0.0186 &3172.26 &-9.933 &0.0134 \\ 
         D &551.07 &9.294 &-0.0142 &3296.22 &-11.769 &0.0164 \\
         H &413.858 &11.441 &-0.0233 &3320.76 &-12.694 &0.0175 \\
         \hline
    \end{tabular}
    \label{tab:melting_quadratic_fit}
\end{table}

\subsection*{Machine-Learning models}

We used message-passing graph neural networks interatomic potentials such as MACE and DPA3 to run large scale MD simulations. In such machine learning models, the potential energy is evaluated as a function of the nuclear positions.
To do so, an input configuration is mapped onto a set of learnable features, which in turn are mapped to the total energy.
The features associated with a given atom depend not only on the coordinates of its neighbors, but also on the neighbor's features.
Feature construction proceeds recursively so that information is propagated to a larger neighborhood. The information transferred from one atom to another is referred to as a message, and each successive feature update constitutes a message-passing step. 

In order to make an estimate of the error resulting from a machine learning approximation we studied the phase diagram using two different models trained on the same QMC dataset.
We chose MACE\cite{batatia2022mace} and DeePMD\cite{zhang2026graphneuralnetworkera} because both were state-of-the-art machine learning models: MACE is a good representative of an equivariant machine learning model and DeePMD is a state-of-the-art representative of an invariant machine learning model\cite{peng2026lambench, kovacs_evaluation_2023}. 
MACE differentiates itself from other equivariant message-passing architectures by explicitly incorporating higher-body-order interactions within the message-passing process. 
The DeePMD model uses invariant features and incorporates higher-body effects by applying line-graph transformation to create higher-order graphs whose vertices correspond to higher-body order terms and then does message passing over that hierarchy~\cite{zhang2025graph}. 

Both models used a loss function of the form:
\begin{equation}
    {\cal L} = \lambda_E \sigma_E^2 + \lambda_F \sigma_F^2
\end{equation}
where
\[
\sigma_E^2=\frac{1}{M}\sum_{s=1}^{M}
\left(\frac{E_s^{\mathrm{ML}}-E_s^{\mathrm{QMC}}}{N}\right)^2
\qquad
\sigma_F^2=\frac{1}{3NM}
\sum_{s=1}^{M}\sum_{i=1}^{N}\sum_{\alpha=x,y,z}
\left(F_{si\alpha}^{\mathrm{ML}}-F_{si\alpha}^{\mathrm{QMC}}\right)^2.
\]
Here, \(E_s\) is the total energy of configuration \(s\), \(N\) is the number of protons in each configuration, and \(M\) is the number of configurations included in the average. Thus, \(\sigma_E\) and \(\sigma_F\) are the RMSEs in energy per atom and in individual Cartesian force components, respectively. The parameter \(\lambda=\lambda_E/\lambda_F\), with units of \(\mathrm{\AA}^{-2}\), controls the relative importance of energy errors versus force errors. The averages are taken over training batches during optimization and over the test set when reporting errors.

\begin{figure}[h]
\centering
\includegraphics[width=0.7\linewidth]{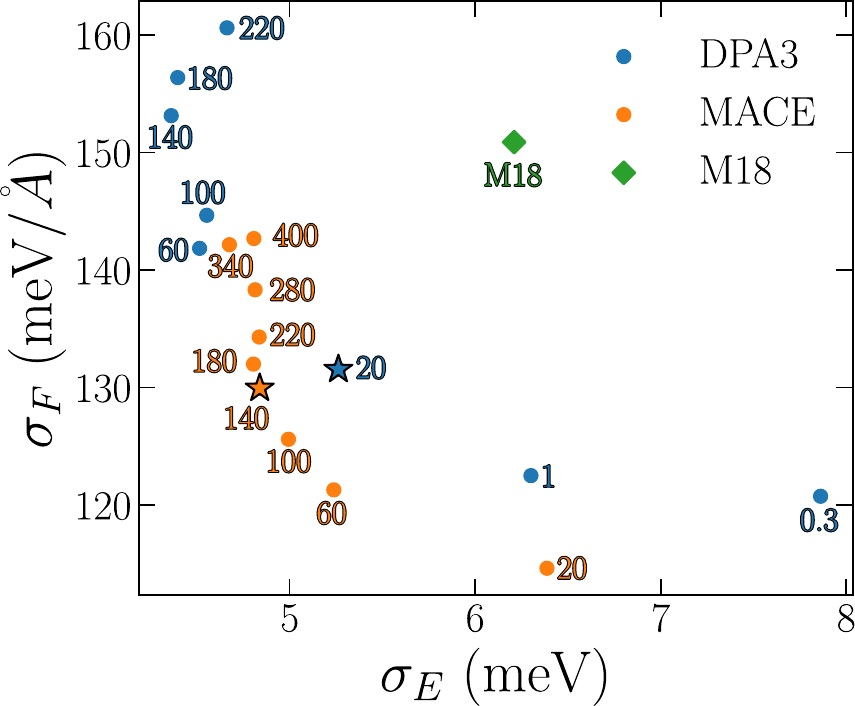}
\caption{
Test-set errors of the MLIP trained on the QMC data. M18 is a reference MACE model from our previous work~\cite{goswami2025high}. The orange and blue stars mark the M29, a MACE model, and \dpm, a DeePMD model, selected to perform the MD simulations. All are trained and tested on the same set of configurations in the region of the LLPT. The numbers shown are $\lambda$ (relative importance of energy over force in the loss function).} 
\label{fig:pareto-M29-DPA3}
\end{figure}

Because the DMC energies are much more accurate than the DMC forces, and because thermodynamic properties are determined by the energy, it is important to increase the importance of the energies  as long as the force RMSE does not increase substantially. 
  Shown in Fig.~\ref{fig:pareto-M29-DPA3} are the errors in the models using various cutoffs and values of $\lambda$.
  We used $\lambda =140$ \(\mathrm{\AA}^{-2}\) for M29 and $\lambda =20 $\(\mathrm{\AA}^{-2}\) for \dpm~ as designated by the stars in fig. \ref{fig:pareto-M29-DPA3}.
  A consideration on the cutoff is given by the size of the simulation box:  for 96 atoms at a density of $0.6/ {\rm \r{A}}^3$,  half the box size is 2.7\r{A}. Having a larger cutoff will result in the interaction between a pair of atoms in certain geometries to be counted twice in the loss function; this has a danger of over-emphasizing periodic effects.
  We settled on cutoffs of $3.5$~ \r{A}~ for M29 and $3.0$~ \r{A}~ for \dpm. 
Our selected MACE model, M29, has a RMSE of energy and force of 4.8 meV/atom and 129.9 meV/\r{A}~ and our best DeePMD model, \dpm, has an average RMSE of 5.3 meV/atom and 131.5 meV/\r{A}.

\begin{figure}[h]
\centering
\includegraphics[width=.7\linewidth]{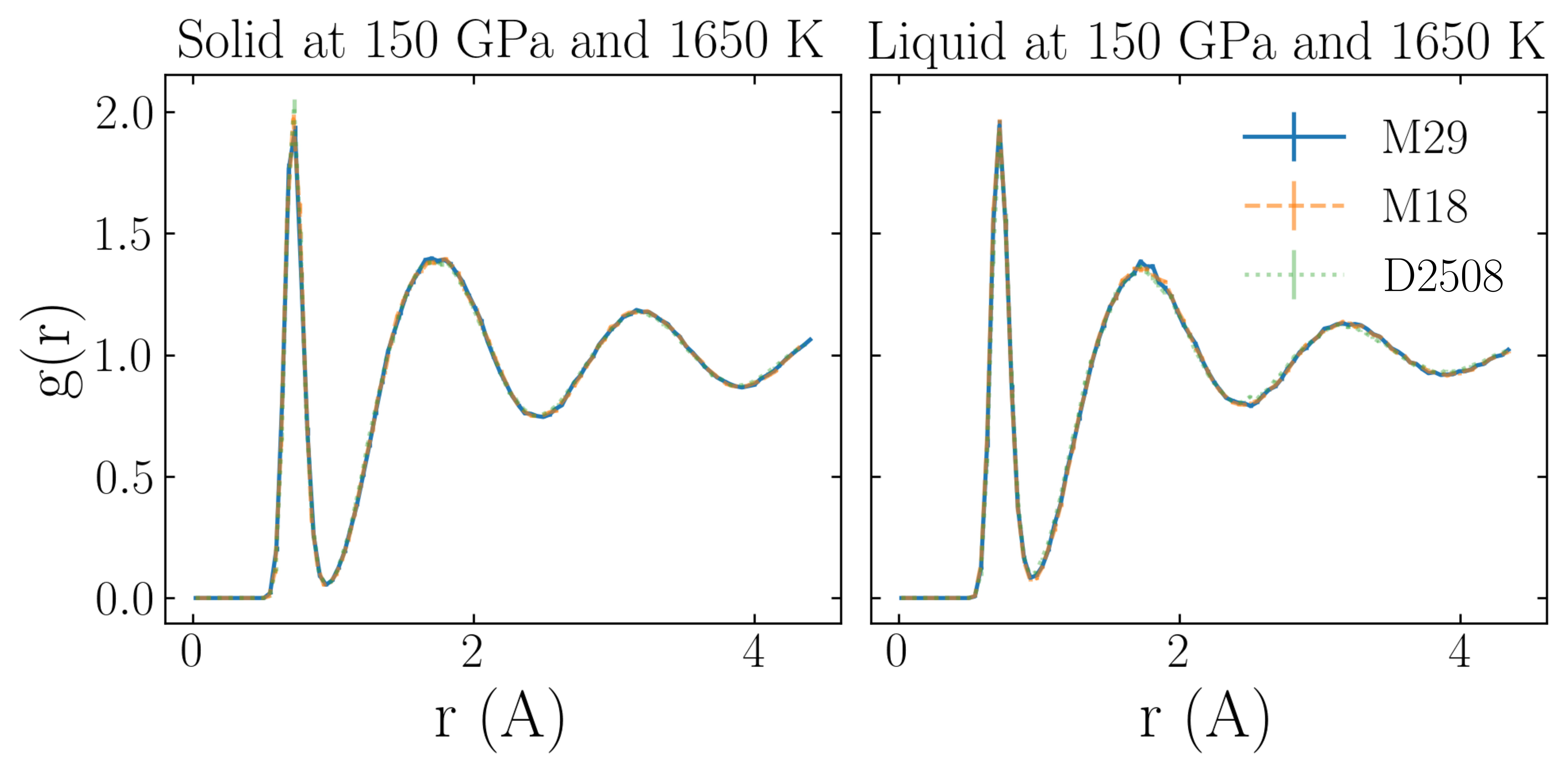}
\caption{Comparison of the radial distribution function between classical protons of M29 and D2508 NPT simulations at 150 GPa for (a) molecular solid (m-HCP) hydrogen at 1650 K and (b) molecular liquid hydrogen at 1650 K.
}
\label{fig:gofr_M18M29DPA3}
\end{figure}

\begin{figure}[h]
    \centering
    \includegraphics[width=0.7\linewidth]{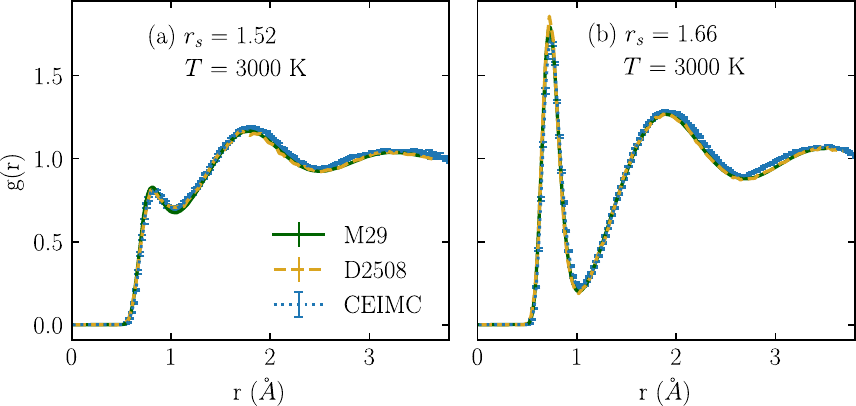}
    \caption{A comparison of the radial distribution functions of quantum protons generated from NVT simulations between \dpm, M29 and CEIMC~\cite{pierleoni2018CPP}. The two models faithfully replicate the atomic and molecular properties of the CEIMC runs, thus generating trust in our models.
    }
    \label{fig:gofr_M29DP2508CEIMC}
\end{figure}

There are no experimental data under these conditions to verify the results obtained from these two models. In our previous paper~\cite{goswami2025high} the M18 model used to predict the melting line of dense molecular hydrogen fits room temperature experimental scattering data quite well.
The new models, M29 and \dpm, are able to replicate the M18 pair correlation function of molecular hydrogen near the melting line as shown in Fig.~\ref{fig:gofr_M18M29DPA3}. 
In addition, our models also match the pair correlation function of Coupled Electron Ion Monte Carlo~\cite{pierleoni2006coupled,pierleoni2016liquid} as seen in Fig.~\ref{fig:gofr_M29DP2508CEIMC} across the molecular dissociation region providing confidence in the models.

\begin{figure}[h]
\centering
\includegraphics[width=0.7\linewidth]{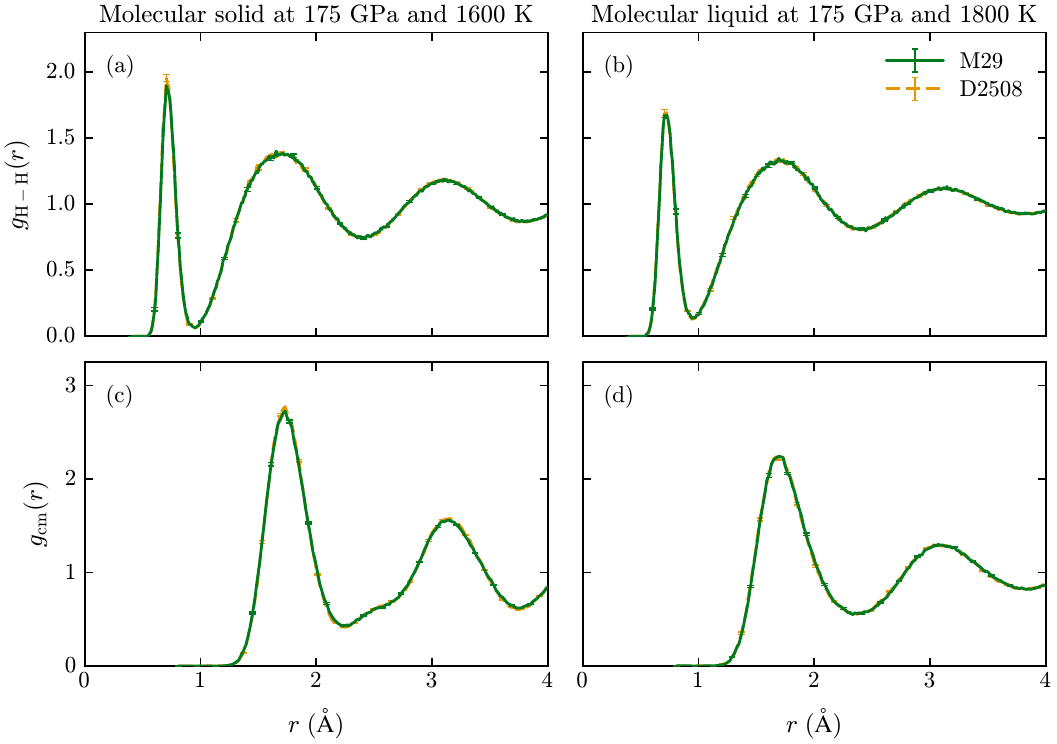}
\caption{
Comparison of the protonic radial distribution functions and that between the molecular centers calculated by classical NVT simulations using M29 and D2508 models for molecular solid hydrogen at 175 GPa and 1600 K and molecular liquid hydrogen at 175 GPa and 1800 K. The upper and lower panels show the proton and molecular-center RDFs, respectively.
}
\label{fig:M29DPA3liquid_solid}
\end{figure}

\DMC{In Fig.~\ref{fig:M29DPA3liquid_solid} we further compare the two models for molecular liquids and molecular solid at $P=175GPa$ near the transition temperature with excellent agreement. }

\subsection*{Enthalpy}
\begin{figure}[h]
    \centering
   \includegraphics[width=0.7\linewidth]{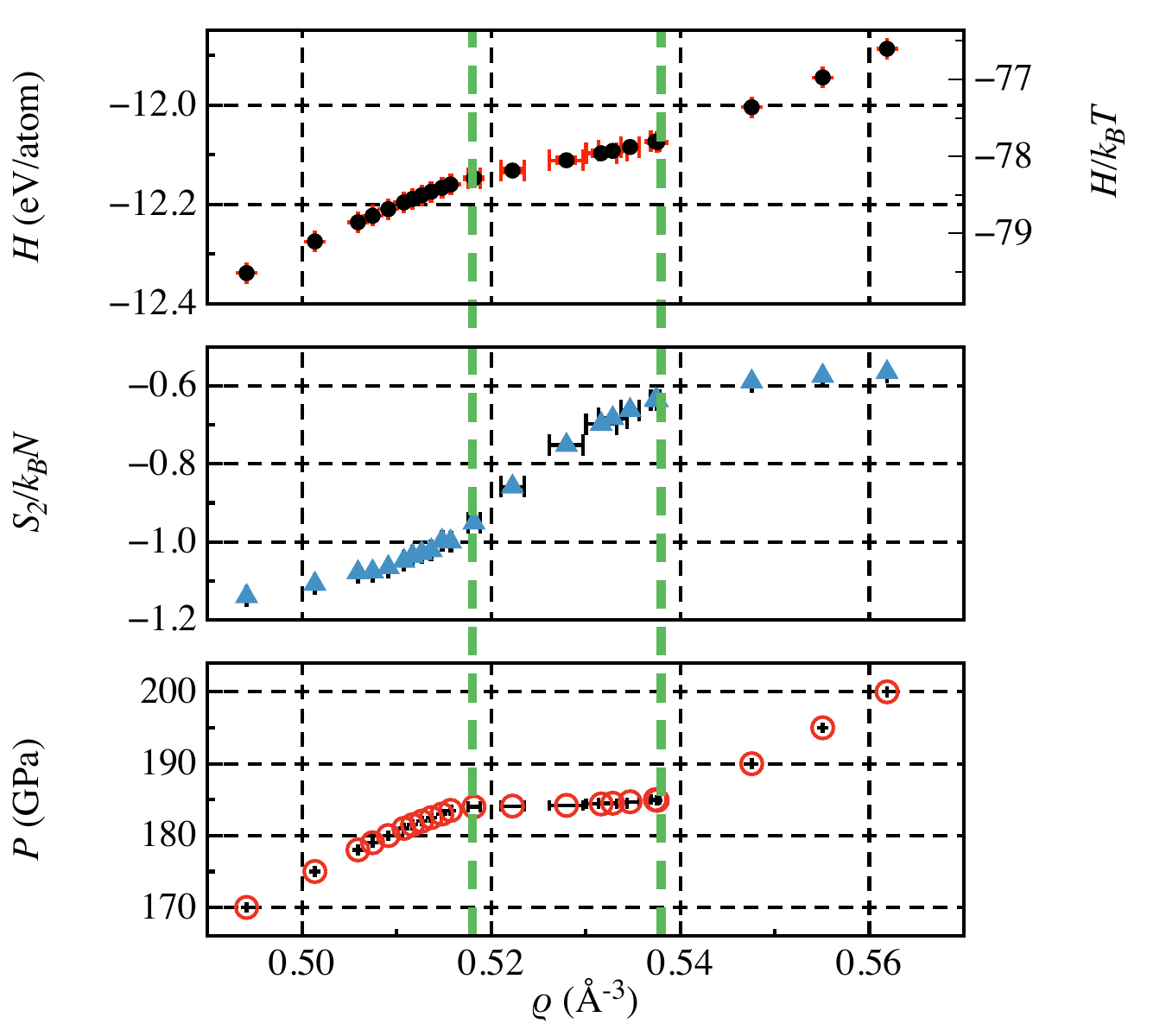}
    \caption{The dependence of enthalpy $H$, pair entropy $S_2$ and pressure $P$ on density at the temperature 1800~K for the classical M29 model.}
    \label{fig:enthalpy}
\end{figure}

In Fig.~\ref{fig:enthalpy}, we show the results of $NPT$ calculations at $T=1800$~K with $N=2592$ atoms. The plateau of the pressure as a function of density indicates the density region where a mixture of the two phases exists. The latent heat can be calculated as a difference of enthalpies $Q = H(\rho_a) - H(\rho_m)$, where $\rho_a$ and $\rho_m$ are atomic and molecular densities at the edges of the two phase region (shaded area in Fig.~\ref{fig:enthalpy}). The dependence of enthalpy $H$ on number density $\rho$ is shown in the upper panel of Fig.~\ref{fig:enthalpy}. The results for $Q$ at temperatures 1700, 1800, 2000, and 2100~K for M29 model are shown in Table~\ref{tab:latent_heat}.
\begin{table}[h]
    \centering
    \caption{Latent heat $Q$ (eV/atom) and pairwise contribution to the latent heat $T\Delta S_2$ (eV/atom) of the LLPT calculated for the classical M29 model of hydrogen. }
    \label{tab:latent_heat}
    \begin{tabular}{|c|c|c|c|c|}
        \hline
         $T$ & $\rho_m$ & $\rho_a$ & $Q$ & $T\Delta S_2$ \\
         \hline
         1700 & 0.531 & 0.550 & 0.071 & 0.048 \\
         1800 & 0.518 & 0.538 & 0.080 & 0.048 \\
         2000 & 0.497 & 0.511 & 0.067 & 0.038 \\
         2100 & 0.491 & 0.500 & 0.040 & 0.023 \\
         \hline
    \end{tabular}
\end{table}

Morales et~al.~\cite{morales2010evidence} calculated the latent heat of phase transition using AIMD with the PBE functional and quantum Monte Carlo to be 0.04--0.05~eV/atom at 1000K. 
Tian et~al.~\cite{tian2019first} estimated the heat of transition at 1000~K using the same method obtaining 0.03--0.04~eV/atom, depending on the size of the unit cell. 
In ref. \cite{pierleoni2006coupled} the latent heat from CEIMC simulations for a system of 54 classical protons was estimated to be $\sim 0.05-0.06 eV/atom$ along the transition line from 600K to 1500K. They estimated that the effect of quantum nuclei was smaller.
In this work, we obtained the latent heat value in the range from 0.04 up to 0.08~eV/atom depending on temperature as shown in Table \ref{tab:latent_heat}. The latent heat decreases while approaching the critical point. 

We can rewrite the latent heat in terms of entropy and expand it as a series of two-particle, three-particle, etc. correlations
\begin{equation}
    Q = T\Delta S = T(\Delta S_2 + \Delta S_3 + \dots).
\end{equation}
The pair entropy $S_2$ is calculated as an integral of the proton-proton pair correlation function $g(r)$ ~\cite{khrapak2021,nicholson2021,gartner2022liquid,nikitiuk2022pair, lukyanchuk2026influence}
\begin{equation}
\frac{S_{2}}{N k_B} = -2\pi\rho \int_{0}^{\infty} \left( g(r) \ln g(r) - g(r) + 1 \right) r^2 dr,
\label{eq:s2_def}
\end{equation}
where $\rho$ is the proton number density. The pairwise entropy contribution to the latent heat can be computed as $T\Delta S_2 = T(S_2(\rho_a) - S_2(\rho_m))$. The results for the dependence of pair entropy $S_2$ on density $\rho$ at $T=1800$~K are shown in the middle panel of Fig.~\ref{fig:enthalpy}. It monotonically grows as the density increases. The rapid increase of pair entropy in the phase transition region points to the structural changes from molecular to atomic liquid. The values of $T\Delta S_{2}$ are presented in Table~\ref{tab:latent_heat}. Thus, we can conclude that the latent heat $Q$ consists of from 60\% to 70\% of the pairwise contribution $T\Delta S_{2}$.

\subsection*{Results for the LLPT for deuterium}
\begin{figure}[h]
    \centering
    \includegraphics[width=0.7\linewidth]{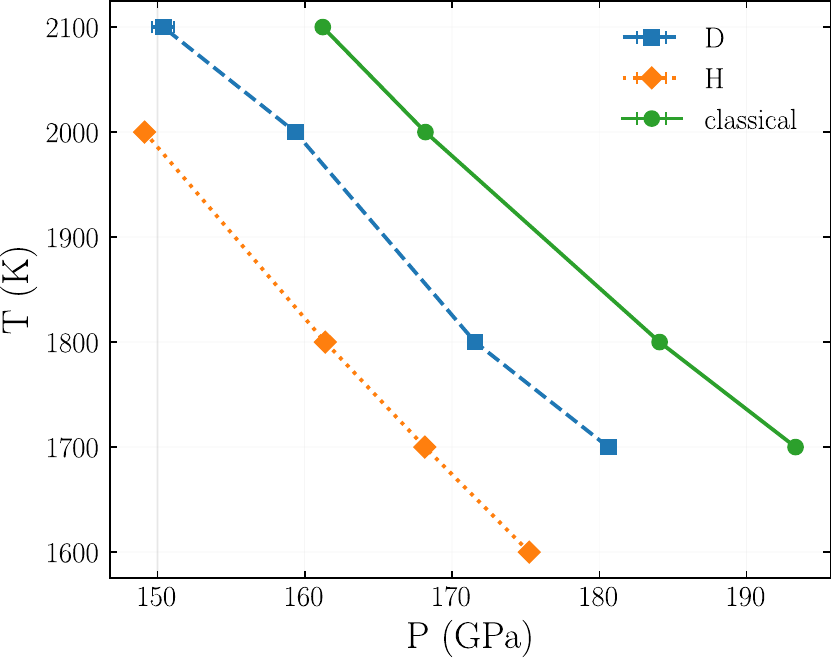}
    \caption{The transition pressures at the LLPT for hydrogen (orange), deuterium (blue) and classical (green) versus temperature. The points were obtained by fitting hyperbolic tangents to the equation of state at each temperature as described in the main text, Eq. (1). }
    \label{fig:deuterium}
\end{figure}

Figure \ref{fig:deuterium} shows PIMD results for deuterium as compared to those of hydrogen and classical protons.  The results for the pressure of the deuterium transition lie close to halfway between  that of classical transition pressure and hydrogen transition pressure.

\subsection*{Derivation of the scaling relation} \label{sec:fss_deriviation}
We derive the dependence of the density with pressure near the critical point shown in the main text
based on known results of the 3D Ising model \cite{binder_finite_nodate} which is in the same universality class.

In the close vicinity of the critical point, we can focus on the
singular part of the free energy in the Ising model for given magnetization
per volume, $s$, in a system of linear extension $L$.
Within the finite size scaling assumption, the behavior of the distribution
function $p_{\rm Ising}^L(s)$ in the immediate vicinity of the critical point 
is then dominated by
\begin{equation}
p_{\rm Ising}^L(s) \sim e^{- \beta F_L^{\rm Ising}(s)} \sim L^x p(s L^y, L/\xi)
\label{fseIsing}
\end{equation}
where $F_L^{\rm Ising}(s)$ is the free energy of the system containing $N=(L/a)^3$ spins
at fixed magnetization $s$, and $a$ is the lattice spacing. For large enough systems, the singular
behavior of the probability distribution is entirely described by
the universal (size independent) scaling function, $p(\cdot, \cdot)$, and in terms of the
correlation length $\xi$.
For the 3D Ising model, one expects $x=y=\beta/ \nu$ and
 $\beta \approx 1/3$, $\nu \approx 2/3$ for the 
critical exponents of the magnetization
$s \sim t^\beta$ and correlation length $\xi \sim t^{-\nu}$ where $t=|T-T_c|/T_c$.

In order to apply the analysis of the Ising model to our case, 
it is convenient to work at fixed pressure
in the $(NPT)$ ensemble, where the
probability distribution for the density $\rho$ of the hydrogen fluid is given by
\begin{equation}    
	p_{\rm fluid}(\rho|P,N,T) \sim 
	\left| \frac{\partial V}{\partial \rho} \right|
	e^{- \beta [F_{\rm fluid}(N,V,T) + P V] } 
	\sim
	\left|\frac{\partial V}{\partial \rho} \right|
	e^{- \beta [F_{\rm fluid}(N,N/\rho,T) + P N/\rho]}.
\end{equation}

Precisely at the critical pressure $P_c$, we can now use the result of the 3D Ising model,
replace the magnetization $s$ by the respective order parameter, the change in density,
$s \to \rho_c^{y/3-1} \Delta \rho \equiv \rho_c^{y/3-1} (\rho-\rho_c)$, where we have inserted $\rho_c^{y/3}$ for dimensional reasons. The singular part of the density distribution
at $P_c$ then writes
\begin{equation}
	p_{\rm fluid}(\rho|P_c,N,T) \sim
	\left|\frac{\partial V}{\partial \rho} \right|
	e^{-\beta [ F_L^{\rm fluid}(N,N/\rho,T)+P_cN/\rho]}
	\sim 
	L^x p( \rho_c^{y/3-1} \Delta \rho L^y, L/\xi)
\end{equation}
where $L = [N/\rho_c]^{1/3}$ characterizes the finite extension of the simulation box at the critical point.

We now introduce $\rho=\rho_c+\Delta \rho$, 
and keep only the leading terms in $\Delta \rho$ in the vicinity of the critical point, e.g.
\begin{equation}
PV \simeq P N [ 1/\rho_c -  \Delta \rho/\rho_c^2 ]
\end{equation}
and drop the Jacobian factor, a constant ($N$) at leading order.
We get
\begin{equation}    
	p_{\rm fluid}(\rho|P,N,T) \sim 
	p_{\rm fluid}(\rho|P_c,N,T) 
	e^{\beta (P-P_c) N \Delta \rho/\rho_c^2 } 
\end{equation}
and we can
study the scaling of the density change
$ \langle \Delta \rho\rangle(N,P,T)$ in the vicinity of $P_c$ and $T_c$
\begin{equation}
	\langle \Delta \rho\rangle(N,P,T) =\frac{\int d (\Delta \rho) \,  p_{\rm fluid}(\rho_c+\Delta \rho|P,N,T)\, \Delta \rho}
	{\int  d (\Delta \rho)  \,  p_{\rm fluid}(\rho_c+\Delta \rho|P,N,T) }
	=
\frac{\int d (\Delta \rho) \, 
\Delta \rho \,
	p(\rho_c^{y/3-1}\Delta \rho L^y,L/\xi)
	e^{\beta (P-P_c)N \Delta \rho/\rho_c^2} }
	{\int  d (\Delta \rho)  \,  p(\rho_c^{y/3-1}\Delta \rho L^y,L/\xi) e^{\beta (P-P_c)N \Delta \rho/\rho_c^2} }.
\end{equation}
Using the change of variables $ u = \rho_c^{y/3-1}\Delta \rho L^{y} $ and $L = \left( \frac{N}{\rho_c}\right)^{-1/3}$
\begin{equation}
 \langle \Delta \rho\rangle \sim \frac{\int d u \, 
u L^{-y} \,
 p(u,L/\xi)
	e^{\beta (P-P_c)N uL^{-y}/\rho_c^2} }
	{\int  d u  \, p(u,L/\xi) e^{\beta (P-P_c)N uL^{-y} \rho_c^{1-y/3}/\rho_c^2} } = L^{-y} f(\beta \Delta P  L^{3-y}\rho_c^{y/3}, L/\xi)
\label{eq:avg_delta_rho_supplementary}
\end{equation}
and $f(\cdot, \cdot)$ denotes the function obtained from the integration, $\Delta P=P_c-P$.
In the main text we used \eqref{eq:avg_delta_rho_supplementary} only at fixed temperature, but one could introduce the scaling relation $\xi \sim \left( (T - T_c) / T_c \right)^{-\nu}$ and work with temperature.

\subsection*{Scaling of $M_0$}
\begin{figure}[h]
    \centering
    \includegraphics[width=0.7\linewidth]{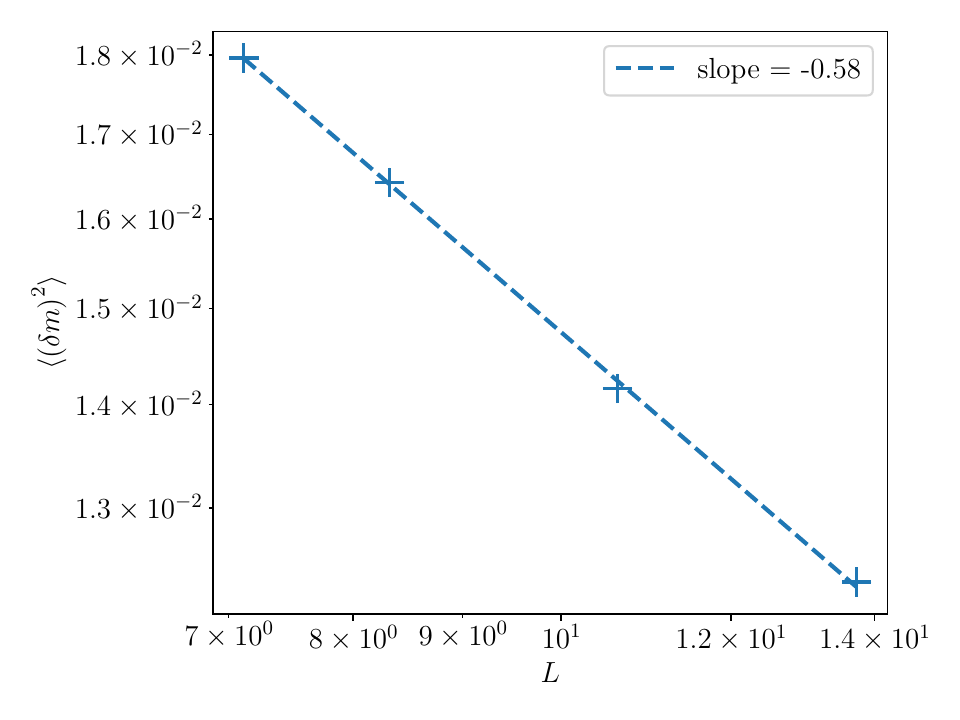}
    \caption{$M_0$ versus $L \sim N^{1/3}$ for estimated $P_c(L)$ and $T_c(L)$ for model M29, in classical simulations, shown with crosses. The straight line illustrates a power law behavior.
    }
    \label{fig:M_0_scaling}
\end{figure}

In the vicinity of the critical point, we verify that $M_0 \sim \langle (\delta m)^2 \rangle$ 
scales as $L^{-\beta / \nu}$ \cite{wilding_simulation_1997}, where $\langle (\delta m)^2 \rangle$ is calculated from the order parameter distribution at criticality, $p_H(\delta m | P_c(L), T_c(L), N)$.
The result for the 4 different system sizes, $N=360, 576, 1200$, and $2592$,
of classical simulations with M29 are shown in fig.~\ref{fig:M_0_scaling},
and show a power law behavior,
$M_0 \propto L^{-c}$, with $c \approx 0.58$, 
in reasonable agreement with the Ising critical exponents estimations, $\beta / \nu \simeq 0.518$.

\subsection*{Estimation of the quantum critical point
}
We estimate the pressure and temperature of the critical point in the quantum and classical case based on the same ansatz for the hydrogen probability distribution.
We derive here the equations for the quantum case, which contains an extra term.
To proceed, we adapt the usual multiple histogram reweighting algorithm \cite{binder_book_monte_carlo, ferrenberg1989optimized, newman1999monte} to path integral molecular dynamics simulations.
We obtain a continuous probability distribution in the variables $(P, T)$.
Compared to usual applications of the method to classical spins systems, we have continuous energies and have to include both the work done by the pressure and the kinetic term $S$, which depends on $\beta^{-1}$.

We first run several simulations in the critical region at different but close pressures and temperatures $(\beta_i, P_i)$.
Configurations are therefore labeled by two integers, one label $s$ for the simulation number running from 1 to $N_s$, and one label $i$ for the $i$th configuration of simulation $s$, running from 1 to $n_s$.
For a simulation $s$ and configuration $i$ sampled at inverse temperature $\beta_k$ and pressure $P_k$ we define
\begin{eqnarray}
H_{\beta_k, P_k}(R_{si}, V_{si}) &\equiv& \beta_k \left(\sum_b^B  \frac{E(R_{si}^b)}{B} + P_kV_{si} \right ) + \frac{S(R_{si})}{\beta_k} \\
S(R_{si}) &=& \sum_b^B \frac{m B}{2 \hbar^2} \left(R_{si}^b - R_{si}^{b+1}\right)^2,
\end{eqnarray}
where $R_{si}^b$ represents all particle positions at bead $b$, $R_{si}$ represents all particle positions at all time slices, and we apply periodic conditions in imaginary time, $R_{si}^{B+1} = R_{si}^1$.
$V_{si}$ is the volume of the simulation box.
The quantity $S(R_{si})$ contains the coupling between beads and $E(R_{si}^b)$ is the potential energy for the time slice $b$, and $B$ is the total number of beads.

At temperatures around $T=2000$K, the average distance between neighboring beads is of the order of $\sqrt{\hbar^2 d / (m k_b T B )} \sim 10^{-4}$~\AA, with $d=3$ the dimension, and $m$ the proton mass.
This is much shorter than interatomic distance, allowing to neglect quantum statistics.
Therefore, the constant pressure PIMD probability distribution for snapshot $i$ of simulation $s$ is \cite{ceperley1995path}
\begin{equation}
    P_{ (N, P_k, T_k)}(R_{si}, V_{si}) = \frac{e^{- H_{\beta_k, P_k}(R_{si}, V_{si}) }}{Z(\beta_k, P_k)}
\end{equation}
where $Z(\beta, P)$ is the partition function. 

To obtain the partition function for each simulation, we solve the equation
\begin{equation}
    Z(\beta_k, P_k) = \sum_s^{N_s} \sum_i^{n_s} \frac{ g_s^{-1}e^{-H_{\beta_k, P_k}(R_{si}, V_{si})} }{  \sum_l^{N_s} \frac{n_l}{g_l} \frac{ e^{- H_{\beta_l, P_l}(R_{si}, V_{si})}}{Z(\beta_l, P_l)}},
\label{eq:partition_func_iter}
\end{equation}  
where $N_s$ is the number of simulations and $n_s$ is the number of configurations from simulation $s$.
$ g_j \equiv 1 + 2 /\tau_j $, with $\tau_j$ the correlation time.
This equation is solved self-consistently starting from a random guess of each $Z(\beta_k, P_k)$ and iterating until convergence.

After \eqref{eq:partition_func_iter} has converged, it is possible to compute $ Z(\beta, P)$ at any pressure and temperature with
\begin{equation}
    Z(\beta, P) = \sum_s^{N_s} \sum_i^{n_s} \frac{ g_s^{-1}e^{-H_{\beta, P}(R_{si}, V_{si})} }{  \sum_l^{N_s} \frac{n_l}{g_l} \frac{ e^{- H_{\beta_l, P_l}(R_{si}, V_{si})}}{Z(\beta_l, P_l)}}.
\end{equation} 

We now compute the density of states $\Omega(E, V, S)$ to estimate the probability distribution for each $(E, V, S)$ possible values
\begin{equation}
    P_{ (N, P, T)}(E, V, S) = \Omega(E, V, S) \frac{e^{-\left( \beta(E + PV) + \frac{S}{\beta} \right) }}{Z(\beta, P)}.
\end{equation}
To do so, we define a 3D grid of 500 points in each direction for $(E, V, S)$ covering the range of values near the critical point.
Knowing $Z(\beta_k, P_k)$, we compute the density of states $\Omega(E, V, S)$ on this grid as 
\begin{equation}
    \Omega(E, V, S) = \frac{\sum_s^{N_s} g_s^{-1} h_s(E, V, S)}{\sum_l^{N_s} \frac{n_l}{g_l} \frac{\exp{\left(-\left( \beta_l(E + P_lV) + S /\beta_l \right) \right )}}{Z(\beta_l, P_l)} }.
\end{equation}
with $h_s(E, V, S)$ the histogram of $(E, V, S)$ values accumulated during simulation $s$.

Finally, to apply the field mixing technique, we sum the distribution on possible kinetic values
\begin{equation}
    \tilde P_{ (N, P, T)}(E, V) = \sum_S P_{ (N, P, T)}(E, V, S).
\end{equation}
From the marginal $\tilde P_{ (N, P, T)}(E, V)$, we then build the new variable $m = N / V + s E / N$ and find the pressure and temperature such that the square difference between $\tilde P_{ (N, P, T)}(m_Q)$ and $p_I(M)$ is minimal.

\subsection*{Comparisons with previous simulations}

Numerous predictions for the LLPT lines are available in the literature since the first FPMD simulations by Scandolo \cite{scandolo2003liquid}. Most of the calculations studied the transition of classical atoms based on PBE forces \cite{Bonev2004,morales2010evidence,Tamblyn2010,lorenzen2010first,karasiev2021liquid} and only provide rough estimates of the critical point,
only Ref.~\cite{istas2025liquid} provided a finite size scaling analysis to locate the critical point within PBE and classical particles based on a ML model (NequIP).
Tenti et al. \cite{tenti_hydrogen_2025} trained a MACE model on PBE to study the Widom line.
The transition line depends rather strongly on the
XC functional used for electronic forces \cite{morales2013,geng2019thermodynamic,Hinz2020,Bund2021,bergermann2024}, typically shifting to higher pressures compared to PBE,
whereas nuclear quantum effects systematically lowers the transition pressures by around $20$GPa for hydrogen compared to the classical line.
Classical MD calculations using VMC forces by Mazzola et al. \cite{Mazzola2018} gave a transition close to CEIMC predictions \cite{pierleoni2016liquid}. Tirelli et al. \cite{tirelli2022high} obtained the same transition line by using $\Delta$-learning of the VMC energies on top of a DFT calculation. 

\begin{figure}[h]
  \centering
   \includegraphics[width=\linewidth]{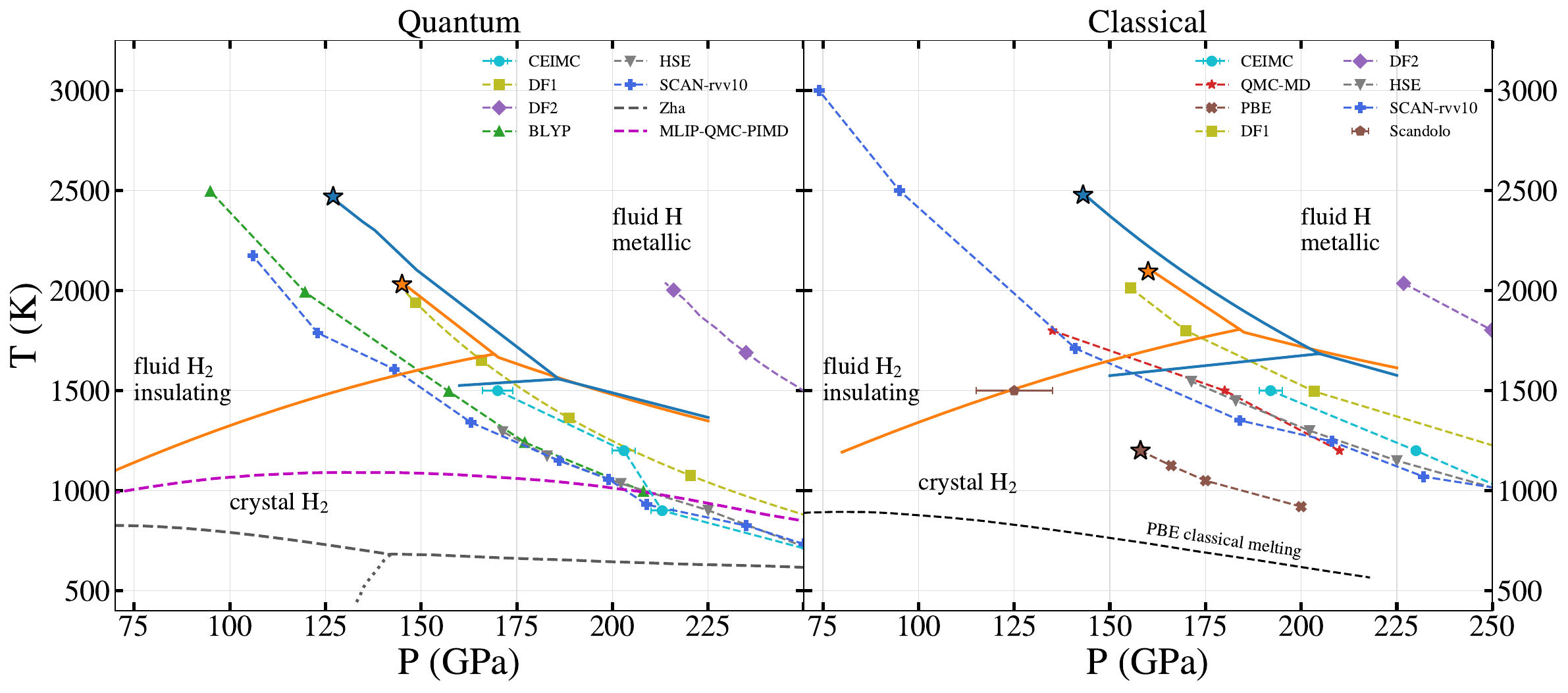}
  \caption{Theoretical phase diagram from different models of quantum hydrogen (left panel) and classical protons (right panel). \DMC{The transition temperatures are indicated by orange line for the M29 model and blue line for the D2508 model.} The LLPT lines for these models terminate at critical points (stars); \DMC{For quantum hydrogen we also show (lower black dashed lines) the experimental melting lines from Rahman spectra (Zha et al. ref. \cite{Zha2017}) and recent PIMD calculations (purple dashed lines) using ML-QMC forces (MLIP-QMC PIMD) (Luo et al. ref. \cite{Luo2026}).}   CEIMC data (cyan closed circle) from ref. \cite{pierleoni2016liquid}, QMC-MD data (red stars) from ref. \cite{Mazzola2018}, PBE results including the location of critical point (brown crosses and large brown star respectively) from ref. \cite{istas2025liquid}, vdW-DF1 data (yellow squares, DF1) and vdW-DF2 (purple diamonds, DF2) from ref.\cite{pierleoni2016liquid}, BLYP data for quantum protons (upward green triangle) from ref. \cite{Bund2021}, HSE data (downward grey triangle) from ref. \cite{bergermann2024}, SCAN-rvv10 data (blue crosses) from ref. \cite{Hinz2020}. The pentagonal brown point in the right panel is the first ab-initio PBE prediction of LLPT by Scandolo \cite{scandolo2003liquid}.
} 
  \label{fig:llpt_theory}
\end{figure}

In Fig. \ref{fig:llpt_theory}, we compare our results to the more recent theoretical predictions involving both classical and quantum protons. As mentioned already at the DFT level of theory, the location of the transition line depends strongly on the adopted XC approximation. As for the location of the critical point we are not aware of any quantitative prediction based on finite-size scaling analysis other than the one in ref. \cite{istas2025liquid} for the DFT-PBE approximation and classical protons.

Most of the previous simulations that employed DFT-XC approximations predict a LLPT line at pressure lower than our present QMC one, except for the vdW-DF2 (DF2 in the figure). The lower pressures can be related to the molecular stability and gap closure of each XC approximation. Also, many XC approximation do not treat long-range dispersion interactions in a consistent way resulting in a lower melting temperature than that estimated using QMC. 
So it is not surprising that many of the DFT calculations located their LLPT line inside our present m-HCP phase. 
It is more surprising that QMC based methods (CEIMC and QMC-DM) also predicted LLPT points inside the crystal phase. As for CEIMC we have checked that this is due to severe nuclear finite-size effects that could not be considered in the original calculations (We could only handle systems of at most 100 atoms for short sampling trajectories). Another anomaly in the CEIMC results is a downward shift of the pressure 
of the order of $20$GPa (probably coming from the simple pressure estimator) with respect to the present ML-QMC models predictions. We plan to investigate in the future the origin of such effect. Comparing our present ML-QMC based predictions with direct calculation once more emphasizes the great importance of learning the most accurate many-body potential energy surface and extend through this methodology the exploration of physical properties to many more atoms combined with a much more exhaustive sampling of the configuration space.

\begin{thebibliography}{67}
\providecommand{\natexlab}[1]{#1}
\providecommand{\url}[1]{\texttt{#1}}
\expandafter\ifx\csname urlstyle\endcsname\relax
  \providecommand{\doi}[1]{doi: #1}\else
  \providecommand{\doi}{doi: \begingroup \urlstyle{rm}\Url}\fi

\bibitem[Wigner and Huntington(1935)]{wigner1935possibility}
E~Wigner and H~B Huntington.
\newblock On the possibility of a metallic modification of hydrogen.
\newblock \emph{The Journal of Chemical Physics}, 3\penalty0 (12):\penalty0
  764--770, 1935.

\bibitem[Mao and Hemley(1994)]{Mao1994}
Ho-kwang Mao and Russell~J. Hemley.
\newblock Ultrahigh-pressure transitions in solid hydrogen.
\newblock \emph{Rev. Mod. Phys.}, 66:\penalty0 671--692, Apr 1994.
\newblock \doi{10.1103/RevModPhys.66.671}.
\newblock URL \url{https://link.aps.org/doi/10.1103/RevModPhys.66.671}.

\bibitem[Ginzburg(1999)]{ginzburg_what_1999}
Vitalii~L. Ginzburg.
\newblock What problems of physics and astrophysics seem now to be especially
  important and interesting (thirty years later, already on the verge of {XXI}
  century)?
\newblock \emph{Phys.-Usp.}, 42\penalty0 (4):\penalty0 353, April 1999.
\newblock ISSN 1063-7869.
\newblock \doi{10.1070/PU1999v042n04ABEH000562}.
\newblock URL
  \url{https://iopscience.iop.org/article/10.1070/PU1999v042n04ABEH000562/meta}.
\newblock Publisher: IOP Publishing.

\bibitem[Zaghoo et~al.(2016)Zaghoo, Salamat, and Silvera]{zaghoo2016evidence}
Mohamed Zaghoo, Ashkan Salamat, and Isaac~F Silvera.
\newblock Evidence of a first-order phase transition to metallic hydrogen.
\newblock \emph{Physical Review B}, 93\penalty0 (15):\penalty0 155128, 2016.
\newblock \doi{10.1103/PhysRevB.93.155128}.

\bibitem[Celliers et~al.(2018)Celliers, Millot, Brygoo, McWilliams,
  Fratanduono, Rygg, Goncharov, Loubeyre, Eggert, Peterson,
  et~al.]{celliers2018insulator}
Peter~M Celliers, Marius Millot, Stephanie Brygoo, R~Stewart McWilliams,
  Dayne~E Fratanduono, J~Ryan Rygg, Alexander~F Goncharov, Paul Loubeyre, Jon~H
  Eggert, J~Luc Peterson, et~al.
\newblock Insulator-metal transition in dense fluid deuterium.
\newblock \emph{Science}, 361\penalty0 (6403):\penalty0 677--682, 2018.
\newblock \doi{DOI: 10.1126/science.aat0970}.

\bibitem[Gregoryanz et~al.(2020)Gregoryanz, Ji, Dalladay-Simpson, Li, Howie,
  and Mao]{gregoryanz2020everything}
Eugene Gregoryanz, Cheng Ji, Philip Dalladay-Simpson, Bing Li, Ross~T Howie,
  and Ho-Kwang Mao.
\newblock Everything you always wanted to know about metallic hydrogen but were
  afraid to ask.
\newblock \emph{Matter and Radiation at Extremes}, 5\penalty0 (3), 2020.

\bibitem[Bonitz et~al.(2024)Bonitz, Vorberger, Bethkenhagen, Böhme, Ceperley,
  Filinov, Gawne, Graziani, Gregori, Hamann, Hansen, Holzmann, Hu, Kählert,
  Karasiev, Kleinschmidt, Kordts, Makait, Militzer, Moldabekov, Pierleoni,
  Preising, Ramakrishna, Redmer, Schwalbe, Svensson, and
  Dornheim]{bonitz_first_2024}
Michael Bonitz, Jan Vorberger, Mandy Bethkenhagen, Maximilian Böhme, David
  Ceperley, Alexey Filinov, Thomas Gawne, Frank Graziani, Gianluca Gregori,
  Paul Hamann, Stephanie Hansen, Markus Holzmann, S.~X. Hu, Hanno Kählert,
  Valentin Karasiev, Uwe Kleinschmidt, Linda Kordts, Christopher Makait,
  Burkhard Militzer, Zhandos Moldabekov, Carlo Pierleoni, Martin Preising,
  Kushal Ramakrishna, Ronald Redmer, Sebastian Schwalbe, Pontus Svensson, and
  Tobias Dornheim.
\newblock Toward first principles simulations of dense hydrogen.
\newblock \emph{Physics of Plasmas}, 31\penalty0 (11):\penalty0 110501,
  November 2024.
\newblock ISSN 1070-664X, 1089-7674.
\newblock \doi{10.1063/5.0219405}.
\newblock URL \url{http://arxiv.org/abs/2405.10627}.
\newblock arXiv:2405.10627 [physics].

\bibitem[Cheng et~al.(2020)Cheng, Mazzola, Pickard, and
  Ceriotti]{cheng2020evidence}
Bingqing Cheng, Guglielmo Mazzola, Chris~J Pickard, and Michele Ceriotti.
\newblock Evidence for supercritical behaviour of high-pressure liquid
  hydrogen.
\newblock \emph{Nature}, 585\penalty0 (7824):\penalty0 217--220, 2020.

\bibitem[Karasiev et~al.(2021)Karasiev, Hinz, Hu, and
  Trickey]{karasiev2021liquid}
Valentin~V Karasiev, Joshua Hinz, SX~Hu, and Samuel~B Trickey.
\newblock On the liquid--liquid phase transition of dense hydrogen.
\newblock \emph{Nature}, 600\penalty0 (7889):\penalty0 E12--E14, 2021.

\bibitem[Tirelli et~al.(2022)Tirelli, Tenti, Nakano, and
  Sorella]{tirelli2022high}
Andrea Tirelli, Giacomo Tenti, Kousuke Nakano, and Sandro Sorella.
\newblock High-pressure hydrogen by machine learning and quantum monte carlo.
\newblock \emph{Physical Review B}, 106\penalty0 (4):\penalty0 L041105, 2022.

\bibitem[Istas et~al.(2025)Istas, Jensen, Yang, Holzmann, Pierleoni, and
  Ceperley]{istas2025liquid}
Mathieu Istas, Scott Jensen, Yubo Yang, Markus Holzmann, Carlo Pierleoni, and
  David~M Ceperley.
\newblock Liquid-liquid phase transition of hydrogen and its critical point:
  Analysis from ab initio simulation and a machine-learned potential.
\newblock \emph{Physical Review E}, 111\penalty0 (4):\penalty0 045307, 2025.

\bibitem[Zel’dovich and Landau(1944)]{Zeldovitch}
Ya. Zel’dovich and L.~Landau.
\newblock On the relation between the liquid and the gaseous states metals.
\newblock \emph{Zhurnal Eksperimentalnoy I Teoreticheskoy Fiziki},
  114:\penalty0 32, 1944.

\bibitem[Scandolo(2003)]{scandolo2003liquid}
Sandro Scandolo.
\newblock Liquid--liquid phase transition in compressed hydrogen from
  first-principles simulations.
\newblock \emph{Proceedings of the National Academy of Sciences}, 100\penalty0
  (6):\penalty0 3051--3053, 2003.

\bibitem[Morales et~al.(2010)Morales, Pierleoni, Schwegler, and
  Ceperley]{morales2010evidence}
Miguel~A Morales, Carlo Pierleoni, Eric Schwegler, and David~M Ceperley.
\newblock Evidence for a first-order liquid-liquid transition in high-pressure
  hydrogen from ab initio simulations.
\newblock \emph{Proceedings of the National Academy of Sciences}, 107\penalty0
  (29):\penalty0 12799--12803, 2010.

\bibitem[Lorenzen et~al.(2010)Lorenzen, Holst, and Redmer]{lorenzen2010first}
Winfried Lorenzen, Bastian Holst, and Ronald Redmer.
\newblock First-order liquid-liquid phase transition in dense hydrogen.
\newblock \emph{Physical Review B—Condensed Matter and Materials Physics},
  82\penalty0 (19):\penalty0 195107, 2010.

\bibitem[Norman and Saitov(2017)]{norman2017critical}
GE~Norman and IM~Saitov.
\newblock Critical point and mechanism of the fluid--fluid phase transition in
  warm dense hydrogen.
\newblock \emph{Doklady physics}, 62:\penalty0 294--298, 2017.
\newblock \doi{https://doi.org/10.1134/S1028335817060088}.

\bibitem[Pierleoni et~al.(2016)Pierleoni, Morales, Rillo, Holzmann, and
  Ceperley]{pierleoni2016liquid}
Carlo Pierleoni, Miguel~A Morales, Giovanni Rillo, Markus Holzmann, and David~M
  Ceperley.
\newblock Liquid--liquid phase transition in hydrogen by coupled electron--ion
  monte carlo simulations.
\newblock \emph{Proceedings of the National Academy of Sciences}, 113\penalty0
  (18):\penalty0 4953--4957, 2016.

\bibitem[Pierleoni et~al.(2018)Pierleoni, Holzmann, and
  Ceperley]{pierleoni2018CPP}
Carlo Pierleoni, Markus Holzmann, and David~M. Ceperley.
\newblock Local structure in dense hydrogen at the liquid–liquid phase
  transition by coupled electron–ion monte carlo.
\newblock \emph{Contributions to Plasma Physics}, 58\penalty0 (2-3):\penalty0
  99--106, 2018.
\newblock \doi{https://doi.org/10.1002/ctpp.201700184}.
\newblock URL
  \url{https://onlinelibrary.wiley.com/doi/abs/10.1002/ctpp.201700184}.

\bibitem[Mazzola et~al.(2018)Mazzola, Helled, and Sorella]{Mazzola2018}
Guglielmo Mazzola, Ravit Helled, and Sandro Sorella.
\newblock {Phase Diagram of Hydrogen and a Hydrogen-Helium Mixture at Planetary
  Conditions by Quantum Monte Carlo Simulations}.
\newblock \emph{Physical Review Letters}, 120:\penalty0 025701, 2018.
\newblock \doi{10.1103/PhysRevLett.120.025701}.
\newblock URL
  \url{https://journals.aps.org/prl/pdf/10.1103/PhysRevLett.120.025701}.

\bibitem[Tenti et~al.(2025)Tenti, Jäckl, Nakano, Rupp, and
  Casula]{tenti_hydrogen_2025}
Giacomo Tenti, Bastian Jäckl, Kousuke Nakano, Matthias Rupp, and Michele
  Casula.
\newblock Hydrogen liquid-liquid transition from first principles and machine
  learning.
\newblock \emph{Physical Review B}, 112\penalty0 (10):\penalty0 104208,
  September 2025.
\newblock ISSN 2469-9950, 2469-9969.
\newblock \doi{10.1103/pbrk-3zgd}.
\newblock URL \url{https://link.aps.org/doi/10.1103/pbrk-3zgd}.

\bibitem[Clay et~al.(2014)Clay, Mcminis, McMahon, Pierleoni, Ceperley, and
  Morales]{Clay2014}
Raymond~C. Clay, Jeremy Mcminis, Jeffrey~M. McMahon, Carlo Pierleoni, David~M.
  Ceperley, and Miguel~A. Morales.
\newblock Benchmarking exchange-correlation functionals for hydrogen at high
  pressures using quantum monte carlo.
\newblock \emph{Phys. Rev. B}, 89:\penalty0 184106, May 2014.
\newblock \doi{10.1103/PhysRevB.89.184106}.
\newblock URL \url{https://link.aps.org/doi/10.1103/PhysRevB.89.184106}.

\bibitem[Niu et~al.(2023)Niu, Yang, Jensen, Holzmann, Pierleoni, Ceperley,
  et~al.]{niu2023stable}
Hongwei Niu, Yubo Yang, Scott Jensen, Markus Holzmann, Carlo Pierleoni, David~M
  Ceperley, et~al.
\newblock Stable solid molecular hydrogen above 900 k from a machine-learned
  potential trained with diffusion quantum monte carlo.
\newblock \emph{Physical review letters}, 130\penalty0 (7):\penalty0 076102,
  2023.

\bibitem[Dzyabura et~al.(2013)Dzyabura, Zaghoo, and
  Silvera]{Dzyabura_PNAS_2013}
Vasily Dzyabura, Mohamed Zaghoo, and Isaac~F. Silvera.
\newblock Evidence of a liquid–liquid phase transition in hot dense hydrogen.
\newblock \emph{Proceedings of the National Academy of Sciences}, 110\penalty0
  (20):\penalty0 8040--8044, 2013.
\newblock \doi{10.1073/pnas.1300718110}.
\newblock URL \url{https://www.pnas.org/doi/abs/10.1073/pnas.1300718110}.

\bibitem[Zaghoo et~al.(2018)Zaghoo, Husband, and Silvera]{zaghoo2018striking}
Mohamed Zaghoo, Rachel~J. Husband, and Isaac~F. Silvera.
\newblock Striking isotope effect on the metallization phase lines of liquid
  hydrogen and deuterium.
\newblock \emph{Phys. Rev. B}, 98:\penalty0 104102, Sep 2018.
\newblock \doi{10.1103/PhysRevB.98.104102}.
\newblock URL \url{https://link.aps.org/doi/10.1103/PhysRevB.98.104102}.

\bibitem[Zha et~al.(2017)Zha, Liu, Tse, and Hemley]{Zha2017}
Chang-Sheng Zha, Hanyu Liu, John~S Tse, and Russell~J Hemley.
\newblock {Melting and High P-T Transitions of Hydrogen up to 300 GPa}.
\newblock \emph{Phys. Rev. Letts.}, 119:\penalty0 075302, 2017.
\newblock \doi{10.1103/PhysRevLett.119.075302}.
\newblock URL
  \url{https://journals.aps.org/prl/pdf/10.1103/PhysRevLett.119.075302}.

\bibitem[Kim et~al.(2018)Kim, Baczewski, Beaudet, Benali, Bennett, Berrill,
  Blunt, Borda, Casula, Ceperley, Chiesa, Clark, Clay, Delaney, Dewing, Esler,
  Hao, Heinonen, Kent, Krogel, Kyl{\"{a}}np{\"{a}}{\"{a}}, Li, Lopez, Luo,
  Malone, Martin, Mathuriya, McMinis, Melton, Mitas, Morales, Neuscamman,
  Parker, {Pineda Flores}, Romero, Rubenstein, Shea, Shin, Shulenburger,
  Tillack, Townsend, Tubman, {Van Der Goetz}, Vincent, Yang, Yang, Zhang, and
  Zhao]{Kim2018}
Jeongnim Kim, Andrew~D Baczewski, Todd~D. Beaudet, Anouar Benali, M.~Chandler
  Bennett, Mark~A. Berrill, Nick~S. Blunt, Edgar Josu{\'{e}}~Landinez Borda,
  Michele Casula, David~M. Ceperley, Simone Chiesa, Bryan~K. Clark, Raymond~C.
  Clay, Kris~T. Delaney, Mark Dewing, Kenneth~P. Esler, Hongxia Hao, Olle
  Heinonen, Paul R~C Kent, Jaron~T. Krogel, Ilkka Kyl{\"{a}}np{\"{a}}{\"{a}},
  Ying~Wai Li, M.~Graham Lopez, Ye~Luo, Fionn~D. Malone, Richard~M. Martin,
  Amrita Mathuriya, Jeremy McMinis, Cody~A. Melton, Lubos Mitas, Miguel~A.
  Morales, Eric Neuscamman, William~D. Parker, Sergio~D. {Pineda Flores},
  Nichols~A. Romero, Brenda~M. Rubenstein, Jacqueline A~R Shea, Hyeondeok Shin,
  Luke Shulenburger, Andreas~F. Tillack, Joshua~P. Townsend, Norm~M. Tubman,
  Brett {Van Der Goetz}, Jordan~E. Vincent, D~ChangMo Yang, Yubo Yang, Shuai
  Zhang, and Luning Zhao.
\newblock {QMCPACK : an open source ab initio quantum Monte Carlo package for
  the electronic structure of atoms, molecules and solids}.
\newblock \emph{Journal of Physics: Condensed Matter}, 30\penalty0
  (19):\penalty0 195901, 5 2018.
\newblock \doi{10.1088/1361-648X/aab9c3}.
\newblock URL
  \url{https://iopscience.iop.org/article/10.1088/1361-648X/aab9c3}.

\bibitem[Kent et~al.(2020)Kent, Annaberdiyev, Benali, Bennett, {Landinez
  Borda}, Doak, Hao, Jordan, Krogel, Kyl{\"{a}}np{\"{a}}{\"{a}}, Lee, Luo,
  Malone, Melton, Mitas, Morales, Neuscamman, Reboredo, Rubenstein, Saritas,
  Upadhyay, Wang, Zhang, and Zhao]{Kent2020}
P.~R.~C. Kent, Abdulgani Annaberdiyev, Anouar Benali, M.~Chandler Bennett,
  Edgar~Josu{\'{e}} {Landinez Borda}, Peter Doak, Hongxia Hao, Kenneth~D.
  Jordan, Jaron~T. Krogel, Ilkka Kyl{\"{a}}np{\"{a}}{\"{a}}, Joonho Lee,
  Ye~Luo, Fionn~D. Malone, Cody~A. Melton, Lubos Mitas, Miguel~A. Morales, Eric
  Neuscamman, Fernando~A. Reboredo, Brenda Rubenstein, Kayahan Saritas, Shiv
  Upadhyay, Guangming Wang, Shuai Zhang, and Luning Zhao.
\newblock {QMCPACK: Advances in the development, efficiency, and application of
  auxiliary field and real-space variational and diffusion quantum Monte
  Carlo}.
\newblock \emph{The Journal of Chemical Physics}, 152\penalty0 (17):\penalty0
  174105, 5 2020.
\newblock ISSN 0021-9606.
\newblock \doi{10.1063/5.0004860}.
\newblock URL \url{https://doi.org/10.1063/5.0004860
  http://aip.scitation.org/doi/10.1063/5.0004860}.

\bibitem[Pierleoni et~al.(2008)Pierleoni, Delaney, Morales, Ceperley, and
  Holzmann]{pierleoni2008}
Carlo Pierleoni, Kris~T. Delaney, Miguel~A. Morales, David~M. Ceperley, and
  Markus Holzmann.
\newblock Trial wave functions for high-pressure metallic hydrogen.
\newblock \emph{Computer Physics Communications}, 179\penalty0 (1):\penalty0
  89--97, 2008.
\newblock ISSN 0010-4655.
\newblock \doi{https://doi.org/10.1016/j.cpc.2008.01.041}.
\newblock URL
  \url{https://www.sciencedirect.com/science/article/pii/S0010465508000210}.
\newblock Special issue based on the Conference on Computational Physics 2007.

\bibitem[Chiesa et~al.(2005)Chiesa, Ceperley, and Zhang]{Chiesa2005}
Simone Chiesa, D.~M. Ceperley, and Shiwei Zhang.
\newblock {Accurate, Efficient, and Simple Forces Computed with Quantum Monte
  Carlo Methods}.
\newblock \emph{Phys. Rev. Lett.}, 94\penalty0 (3):\penalty0 036404, 1 2005.
\newblock ISSN 0031-9007.
\newblock \doi{10.1103/PhysRevLett.94.036404}.
\newblock URL \url{https://link.aps.org/doi/10.1103/PhysRevLett.94.036404}.

\bibitem[Batatia et~al.(2022)Batatia, Kovacs, Simm, Ortner, and
  Cs{\'a}nyi]{batatia2022mace}
Ilyes Batatia, David~P Kovacs, Gregor Simm, Christoph Ortner, and G{\'a}bor
  Cs{\'a}nyi.
\newblock Mace: Higher order equivariant message passing neural networks for
  fast and accurate force fields.
\newblock \emph{Advances in Neural Information Processing Systems},
  35:\penalty0 11423--11436, 2022.

\bibitem[Zhang et~al.(2026)Zhang, Peng, Cai, Li, Zhou, Zeng, Guo, Zhang, Li,
  Jiang, Zhu, Jia, Zhang, and Wang]{zhang2026graphneuralnetworkera}
Duo Zhang, Anyang Peng, Chun Cai, Wentao Li, Yuanchang Zhou, Jinzhe Zeng,
  Mingyu Guo, Chengqian Zhang, Bowen Li, Hong Jiang, Tong Zhu, Weile Jia,
  Linfeng Zhang, and Han Wang.
\newblock A graph neural network for the era of large atomistic models.
\newblock \emph{arXiv preprint arXiv:2506.01686}, 2026.
\newblock URL \url{https://arxiv.org/abs/2506.01686}.

\bibitem[Ceperley et~al.(2023)Ceperley, Jensen, Yang, Niu, Pierleoni, and
  Holzmann]{ceperley2023training}
David~M Ceperley, Scott Jensen, Yubo Yang, Hongwei Niu, Carlo Pierleoni, and
  Markus Holzmann.
\newblock Training models using forces computed by stochastic electronic
  structure methods.
\newblock \emph{arXiv preprint arXiv:2310.15994}, 2023.

\bibitem[Peng et~al.(2026)Peng, Cai, Guo, Zhang, Zhang, Jiang, Wang, Loew, Wu,
  E, et~al.]{peng2026lambench}
Anyang Peng, Chun Cai, Mingyu Guo, Duo Zhang, Chengqian Zhang, Wanrun Jiang,
  Yinan Wang, Antoine Loew, Chengkun Wu, Weinan E, et~al.
\newblock Lambench: a benchmark for large atomistic models.
\newblock \emph{npj Computational Materials}, 2026.

\bibitem[Kovács et~al.(2023)Kovács, Batatia, Arany, and
  Csányi]{kovacs_evaluation_2023}
Dávid~Péter Kovács, Ilyes Batatia, Eszter~Sára Arany, and Gábor Csányi.
\newblock Evaluation of the {MACE} force field architecture: {From} medicinal
  chemistry to materials science.
\newblock \emph{The Journal of Chemical Physics}, 159\penalty0 (4):\penalty0
  044118, July 2023.
\newblock ISSN 0021-9606, 1089-7690.
\newblock \doi{10.1063/5.0155322}.
\newblock URL
  \url{https://pubs.aip.org/jcp/article/159/4/044118/2904837/Evaluation-of-the-MACE-force-field-architecture}.

\bibitem[Zhang et~al.(2025)Zhang, Peng, Cai, Li, Zhou, Zeng, Guo, Zhang, Li,
  Jiang, et~al.]{zhang2025graph}
Duo Zhang, Anyang Peng, Chun Cai, Wentao Li, Yuanchang Zhou, Jinzhe Zeng,
  Mingyu Guo, Chengqian Zhang, Bowen Li, Hong Jiang, et~al.
\newblock Graph neural network model for the era of large atomistic models.
\newblock \emph{arXiv preprint arXiv:2506.01686}, 2025.

\bibitem[Thompson et~al.(2022)Thompson, Aktulga, Berger, Bolintineanu, Brown,
  Crozier, in~'t Veld, Kohlmeyer, Moore, Nguyen, Shan, Stevens, Tranchida,
  Trott, and Plimpton]{LAMMPS}
A.~P. Thompson, H.~M. Aktulga, R.~Berger, D.~S. Bolintineanu, W.~M. Brown,
  P.~S. Crozier, P.~J. in~'t Veld, A.~Kohlmeyer, S.~G. Moore, T.~D. Nguyen,
  R.~Shan, M.~J. Stevens, J.~Tranchida, C.~Trott, and S.~J. Plimpton.
\newblock {LAMMPS} - a flexible simulation tool for particle-based materials
  modeling at the atomic, meso, and continuum scales.
\newblock \emph{Comp. Phys. Comm.}, 271:\penalty0 108171, 2022.
\newblock \doi{10.1016/j.cpc.2021.108171}.

\bibitem[Li et~al.(2026)Li, Gomez, Cai, Zhang, Fu, Jia, Feldman, Blumer, Higer,
  Hirshberg, et~al.]{li2026fix}
Yifan Li, Axel Gomez, Kehan Cai, Chunyi Zhang, Li~Fu, Weile Jia, Yotam~MY
  Feldman, Ofir Blumer, Jacob Higer, Barak Hirshberg, et~al.
\newblock fix pimd/langevin: An efficient implementation of path integral
  molecular dynamics in lammps.
\newblock \emph{The Journal of Chemical Physics}, 164\penalty0 (14), 2026.

\bibitem[Litman et~al.(2024)Litman, Kapil, Feldman, Tisi, Begušić, Fidanyan,
  Fraux, Higer, Kellner, Li, Pós, Stocco, Trenins, Hirshberg, Rossi, and
  Ceriotti]{iPI3}
Yair Litman, Venkat Kapil, Yotam M.~Y. Feldman, Davide Tisi, Tomislav
  Begušić, Karen Fidanyan, Guillaume Fraux, Jacob Higer, Matthias Kellner,
  Tao~E. Li, Eszter~S. Pós, Elia Stocco, George Trenins, Barak Hirshberg,
  Mariana Rossi, and Michele Ceriotti.
\newblock i-pi 3.0: A flexible and efficient framework for advanced atomistic
  simulations.
\newblock \emph{The Journal of Chemical Physics}, 161\penalty0 (6):\penalty0
  062504, 08 2024.
\newblock ISSN 0021-9606.
\newblock \doi{10.1063/5.0215869}.
\newblock URL \url{https://doi.org/10.1063/5.0215869}.

\bibitem[Goswami et~al.(2025)Goswami, Jensen, Yang, Holzmann, Pierleoni, and
  Ceperley]{goswami2025high}
Shubhang Goswami, Scott Jensen, Yubo Yang, Markus Holzmann, Carlo Pierleoni,
  and David~M Ceperley.
\newblock High temperature melting of dense molecular hydrogen from
  machine-learning interatomic potentials trained on quantum monte carlo.
\newblock \emph{The Journal of Chemical Physics}, 162\penalty0 (5), 2025.

\bibitem[Bonev et~al.(2004)Bonev, Militzer, and Galli]{Bonev2004}
Stanimir~A. Bonev, Burkhard Militzer, and Giulia Galli.
\newblock Ab initio simulations of dense liquid deuterium: Comparison with
  gas-gun shock-wave experiments.
\newblock \emph{Phys. Rev. B}, 69:\penalty0 014101, Jan 2004.
\newblock \doi{10.1103/PhysRevB.69.014101}.
\newblock URL \url{https://link.aps.org/doi/10.1103/PhysRevB.69.014101}.

\bibitem[Luo et~al.(2026)Luo, Ichibha, Ma, Liu, Li, Maezono, Zhong, Xie, Liu,
  Hemley, and Ma]{Luo2026}
Jie Luo, Tom Ichibha, Yuan Ma, Yuan Liu, Jiaxiang Li, Ryo Maezono, Xin Zhong,
  Yu~Xie, Hanyu Liu, Russell~J. Hemley, and Yanming Ma.
\newblock Quantum melting behavior of dense solid hydrogen from large-scale
  machine-learned potential simulations.
\newblock \emph{Phys. Rev. B}, 113:\penalty0 184116, May 2026.
\newblock \doi{10.1103/47cf-cj2s}.
\newblock URL \url{https://link.aps.org/doi/10.1103/47cf-cj2s}.

\bibitem[Binder and Landau(1984)]{binder_finite_size_1984}
K.~Binder and D.~P. Landau.
\newblock Finite-size scaling at first-order phase transitions.
\newblock \emph{Physical Review B}, 30\penalty0 (3):\penalty0 1477--1485,
  August 1984.
\newblock ISSN 0163-1829.
\newblock \doi{10.1103/PhysRevB.30.1477}.
\newblock URL \url{https://link.aps.org/doi/10.1103/PhysRevB.30.1477}.

\bibitem[Binder et~al.(2019)Binder, Heermann, and
  Binder]{binder_book_monte_carlo}
Kurt Binder, Dieter~W Heermann, and K~Binder.
\newblock \emph{Monte Carlo simulation in statistical physics}, volume~8.
\newblock Springer, 2019.

\bibitem[Ferrenberg and Swendsen(1989)]{ferrenberg1989optimized}
Alan~M Ferrenberg and Robert~H Swendsen.
\newblock Optimized monte carlo data analysis.
\newblock \emph{Physical review letters}, 63\penalty0 (12):\penalty0 1195,
  1989.

\bibitem[Binder(1981)]{binder_finite_nodate}
K~Binder.
\newblock Finite size scaling analysis of ising model block distribution
  functions.
\newblock \emph{Zeitschrift fur Physik B Condensed Matter}, 43\penalty0
  (2):\penalty0 119--140, 1981.

\bibitem[Wilding(1997)]{wilding_simulation_1997}
Nigel~B Wilding.
\newblock Simulation studies of fluid critical behaviour.
\newblock \emph{Journal of Physics: Condensed Matter}, 9\penalty0 (3):\penalty0
  585--612, 1 1997.
\newblock ISSN 0953-8984, 1361-648X.
\newblock \doi{10.1088/0953-8984/9/3/002}.
\newblock URL
  \url{https://iopscience.iop.org/article/10.1088/0953-8984/9/3/002}.

\bibitem[Wilding and Binder(1996)]{wilding_finite-size_1996}
N.B. Wilding and K.~Binder.
\newblock Finite-size scaling for near-critical continuum fluids at constant
  pressure.
\newblock \emph{Physica A: Statistical Mechanics and its Applications},
  231\penalty0 (4):\penalty0 439--447, October 1996.
\newblock ISSN 03784371.
\newblock \doi{10.1016/0378-4371(96)00213-0}.
\newblock URL
  \url{https://linkinghub.elsevier.com/retrieve/pii/0378437196002130}.

\bibitem[Debenedetti et~al.(2020)Debenedetti, Sciortino, and
  Zerze]{debenedetti_second_2020}
Pablo~G. Debenedetti, Francesco Sciortino, and Gül~H. Zerze.
\newblock Second critical point in two realistic models of water.
\newblock \emph{Science}, 369\penalty0 (6501):\penalty0 289--292, July 2020.
\newblock ISSN 0036-8075, 1095-9203.
\newblock \doi{10.1126/science.abb9796}.
\newblock URL \url{https://www.science.org/doi/10.1126/science.abb9796}.

\bibitem[Tsypin and Blöte(2000)]{tsypin_probability_2000}
M.~M. Tsypin and H.~W.~J. Blöte.
\newblock Probability distribution of the order parameter for the
  three-dimensional {Ising}-model universality class: {A} high-precision
  {Monte} {Carlo} study.
\newblock \emph{Physical Review E}, 62\penalty0 (1):\penalty0 73--76, July
  2000.
\newblock ISSN 1063-651X, 1095-3787.
\newblock \doi{10.1103/PhysRevE.62.73}.
\newblock URL \url{https://link.aps.org/doi/10.1103/PhysRevE.62.73}.

\bibitem[Ferrenberg and Landau(1991)]{ferrenberg1991critical}
Alan~M. Ferrenberg and D.~P. Landau.
\newblock Critical behavior of the three-dimensional ising model: A
  high-resolution monte carlo study.
\newblock \emph{Phys. Rev. B}, 44:\penalty0 5081--5091, Sep 1991.
\newblock \doi{10.1103/PhysRevB.44.5081}.
\newblock URL \url{https://link.aps.org/doi/10.1103/PhysRevB.44.5081}.

\bibitem[Newman and Barkema(1999)]{newman1999monte}
Mark Newman and Gerard Barkema.
\newblock Monte carlo methods in statistical physics chapter 1-4.
\newblock \emph{New York, USA}, 1999.

\bibitem[Chen et~al.(1982)Chen, Fisher, and Nickel]{chen1982unbiased}
Jing-Huei Chen, Michael~E Fisher, and Bernie~G Nickel.
\newblock Unbiased estimation of corrections to scaling by partial differential
  approximants.
\newblock \emph{Physical Review Letters}, 48\penalty0 (9):\penalty0 630, 1982.

\bibitem[Wilding(1995)]{wilding1995critical}
Nigel~B Wilding.
\newblock Critical-point and coexistence-curve properties of the lennard-jones
  fluid: A finite-size scaling study.
\newblock \emph{Physical Review E}, 52\penalty0 (1):\penalty0 602, 1995.

\bibitem[Tamblyn and Bonev(2010)]{Tamblyn2010}
Isaac Tamblyn and Stanimir~A. Bonev.
\newblock Structure and phase boundaries of compressed liquid hydrogen.
\newblock \emph{Phys. Rev. Lett.}, 104:\penalty0 065702, Feb 2010.
\newblock \doi{10.1103/PhysRevLett.104.065702}.
\newblock URL \url{https://link.aps.org/doi/10.1103/PhysRevLett.104.065702}.

\bibitem[Morales et~al.(2013)Morales, McMahon, Pierleoni, and
  Ceperley]{Morales2013}
Miguel~A. Morales, Jeffrey~M. McMahon, Carlo Pierleoni, and David~M. Ceperley.
\newblock Nuclear quantum effects and nonlocal exchange-correlation functionals
  applied to liquid hydrogen at high pressure.
\newblock \emph{Phys. Rev. Lett.}, 110:\penalty0 065702, Feb 2013.
\newblock \doi{10.1103/PhysRevLett.110.065702}.
\newblock URL \url{https://link.aps.org/doi/10.1103/PhysRevLett.110.065702}.

\bibitem[Geng et~al.(2019)Geng, Wu, Marqu{\'e}s, and
  Ackland]{geng2019thermodynamic}
Hua~Y Geng, Q~Wu, Miriam Marqu{\'e}s, and Graeme~J Ackland.
\newblock Thermodynamic anomalies and three distinct liquid-liquid transitions
  in warm dense liquid hydrogen.
\newblock \emph{Physical Review B}, 100\penalty0 (13):\penalty0 134109, 2019.

\bibitem[Hinz et~al.(2020)Hinz, Karasiev, Hu, Zaghoo,
  Mej\'{\i}a-Rodr\'{\i}guez, Trickey, and Calder\'{\i}n]{Hinz2020}
Joshua Hinz, Valentin~V. Karasiev, S.~X. Hu, Mohamed Zaghoo, Daniel
  Mej\'{\i}a-Rodr\'{\i}guez, S.~B. Trickey, and L.~Calder\'{\i}n.
\newblock Fully consistent density functional theory determination of the
  insulator-metal transition boundary in warm dense hydrogen.
\newblock \emph{Phys. Rev. Res.}, 2:\penalty0 032065(R), Sep 2020.
\newblock \doi{10.1103/PhysRevResearch.2.032065}.
\newblock URL \url{https://link.aps.org/doi/10.1103/PhysRevResearch.2.032065}.

\bibitem[van~de Bund et~al.(2021)van~de Bund, Wiebe, and Ackland]{Bund2021}
Sebastiaan van~de Bund, Heather Wiebe, and Graeme~J. Ackland.
\newblock Isotope quantum effects in the metallization transition in liquid
  hydrogen.
\newblock \emph{Phys. Rev. Lett.}, 126:\penalty0 225701, Jun 2021.
\newblock \doi{10.1103/PhysRevLett.126.225701}.

\bibitem[Bergermann et~al.(2024)Bergermann, Kleindienst, and
  Redmer]{bergermann2024}
Armin Bergermann, Lucas Kleindienst, and Ronald Redmer.
\newblock Nonmetal-to-metal transition in liquid hydrogen using density
  functional theory and the heyd–scuseria–ernzerhof exchange-correlation
  functional.
\newblock \emph{The Journal of Chemical Physics}, 161\penalty0 (23):\penalty0
  234303, 12 2024.
\newblock ISSN 0021-9606.
\newblock \doi{10.1063/5.0241111}.
\newblock URL \url{https://doi.org/10.1063/5.0241111}.

\bibitem[Weir et~al.(1996)Weir, Mitchell, and Nellis]{weir1996metallization}
ST~Weir, AC~Mitchell, and William~J Nellis.
\newblock Metallization of fluid molecular hydrogen at 140 gpa (1.4 mbar).
\newblock \emph{Physical review letters}, 76\penalty0 (11):\penalty0 1860,
  1996.
\newblock \doi{https://doi.org/10.1103/PhysRevLett.76.1860}.

\bibitem[Fortov et~al.(2007)Fortov, Ilkaev, Arinin, Burtzev, Golubev,
  Iosilevskiy, Khrustalev, Mikhailov, Mochalov, Ternovoi,
  et~al.]{fortov2007phase}
VE~Fortov, RI~Ilkaev, VA~Arinin, VV~Burtzev, VA~Golubev, IL~Iosilevskiy,
  VV~Khrustalev, AL~Mikhailov, MA~Mochalov, V~Ya Ternovoi, et~al.
\newblock Phase transition in a strongly nonideal deuterium plasma generated by
  quasi-isentropical compression at megabar pressures.
\newblock \emph{Physical Review Letters}, 99\penalty0 (18):\penalty0 185001,
  2007.
\newblock \doi{https://doi.org/10.1103/PhysRevLett.99.185001}.

\bibitem[Mochalov et~al.(2017)Mochalov, Il’Kaev, Fortov, Mikhailov, Blikov,
  Ogorodnikov, Gryaznov, and Iosilevskii]{mochalov2017quasi}
MA~Mochalov, RI~Il’Kaev, VE~Fortov, AL~Mikhailov, AO~Blikov, VA~Ogorodnikov,
  VK~Gryaznov, and IL~Iosilevskii.
\newblock Quasi-isentropic compressibility of a strongly nonideal deuterium
  plasma at pressures of up to 5500 gpa: Nonideality and degeneracy effects.
\newblock \emph{Journal of Experimental and Theoretical Physics}, 124:\penalty0
  505--529, 2017.

\bibitem[Knudson et~al.(2015)Knudson, Desjarlais, Becker, Lemke, Cochrane,
  Savage, Bliss, Mattsson, and Redmer]{knudson2015direct}
Marcus~D Knudson, Michael~P Desjarlais, Andeas Becker, Raymond~W Lemke,
  KR~Cochrane, Mark~E Savage, David~E Bliss, TR~Mattsson, and Ronald Redmer.
\newblock Direct observation of an abrupt insulator-to-metal transition in
  dense liquid deuterium.
\newblock \emph{Science}, 348\penalty0 (6242):\penalty0 1455--1460, 2015.

\bibitem[Jiang et~al.(2020)Jiang, Holtgrewe, Geballe, Lobanov, Mahmood,
  McWilliams, and Goncharov]{Jiang2020}
Shuqing Jiang, Nicholas Holtgrewe, Zachary~M. Geballe, Sergey~S. Lobanov,
  Mohammad~F. Mahmood, R.~Stewart McWilliams, and Alexander~F. Goncharov.
\newblock {A Spectroscopic Study of the Insulator–Metal Transition in Liquid
  Hydrogen and Deuterium}.
\newblock \emph{Advanced Science}, 7\penalty0 (2), 2020.
\newblock ISSN 21983844.
\newblock \doi{10.1002/advs.201901668}.

\bibitem[Rillo et~al.(2019)Rillo, Morales, Ceperley, and Pierleoni]{Rillo2019}
Giovanni Rillo, Miguel~A. Morales, David~M. Ceperley, and Carlo Pierleoni.
\newblock Optical properties of high-pressure fluid hydrogen across molecular
  dissociation.
\newblock \emph{Proceedings of the National Academy of Sciences}, 116\penalty0
  (20):\penalty0 9770--9774, 2019.
\newblock \doi{10.1073/pnas.1818897116}.
\newblock URL \url{https://www.pnas.org/doi/abs/10.1073/pnas.1818897116}.

\bibitem[Subramanian et~al.(2011)Subramanian, Goncharov, Struzhkin, Somayazulu,
  and Hemley]{Subramanian2011}
Natarajan Subramanian, Alexander~F. Goncharov, Viktor~V. Struzhkin, Maddury
  Somayazulu, and Russell~J. Hemley.
\newblock Bonding changes in hot fluid hydrogen at megabar pressures.
\newblock \emph{Proceedings of the National Academy of Sciences}, 108\penalty0
  (15):\penalty0 6014--6019, 2011.
\newblock \doi{10.1073/pnas.1102760108}.
\newblock URL \url{https://www.pnas.org/doi/abs/10.1073/pnas.1102760108}.

\bibitem[Boerner et~al.(2023)Boerner, Deems, Furlani, Knuth, and
  Towns]{NSF_ACCESS_reference}
Timothy~J. Boerner, Stephen Deems, Thomas~R. Furlani, Shelley~L. Knuth, and
  John Towns.
\newblock Access: Advancing innovation: Nsf’s advanced cyberinfrastructure
  coordination ecosystem: Services \& support.
\newblock In \emph{Practice and Experience in Advanced Research Computing 2023:
  Computing for the Common Good}, PEARC '23, page 173–176, New York, NY, USA,
  2023. Association for Computing Machinery.
\newblock ISBN 9781450399852.
\newblock \doi{10.1145/3569951.3597559}.
\newblock URL \url{https://doi.org/10.1145/3569951.3597559}.

\end{thebibliography}

\begin{thebibliography}{48}
\providecommand{\natexlab}[1]{#1}
\providecommand{\url}[1]{\texttt{#1}}
\expandafter\ifx\csname urlstyle\endcsname\relax
  \providecommand{\doi}[1]{doi: #1}\else
  \providecommand{\doi}{doi: \begingroup \urlstyle{rm}\Url}\fi

\bibitem[Niu et~al.(2023)Niu, Yang, Jensen, Holzmann, Pierleoni, Ceperley,
  et~al.]{niu2023stable}
Hongwei Niu, Yubo Yang, Scott Jensen, Markus Holzmann, Carlo Pierleoni, David~M
  Ceperley, et~al.
\newblock Stable solid molecular hydrogen above 900 k from a machine-learned
  potential trained with diffusion quantum monte carlo.
\newblock \emph{Physical review letters}, 130\penalty0 (7):\penalty0 076102,
  2023.

\bibitem[Goswami et~al.(2025)Goswami, Jensen, Yang, Holzmann, Pierleoni, and
  Ceperley]{goswami2025high}
Shubhang Goswami, Scott Jensen, Yubo Yang, Markus Holzmann, Carlo Pierleoni,
  and David~M Ceperley.
\newblock High temperature melting of dense molecular hydrogen from
  machine-learning interatomic potentials trained on quantum monte carlo.
\newblock \emph{The Journal of Chemical Physics}, 162\penalty0 (5), 2025.

\bibitem[Kim et~al.(2018)Kim, Baczewski, Beaudet, Benali, Bennett, Berrill,
  Blunt, Borda, Casula, Ceperley, Chiesa, Clark, Clay, Delaney, Dewing, Esler,
  Hao, Heinonen, Kent, Krogel, Kyl{\"{a}}np{\"{a}}{\"{a}}, Li, Lopez, Luo,
  Malone, Martin, Mathuriya, McMinis, Melton, Mitas, Morales, Neuscamman,
  Parker, {Pineda Flores}, Romero, Rubenstein, Shea, Shin, Shulenburger,
  Tillack, Townsend, Tubman, {Van Der Goetz}, Vincent, Yang, Yang, Zhang, and
  Zhao]{Kim2018}
Jeongnim Kim, Andrew~D Baczewski, Todd~D. Beaudet, Anouar Benali, M.~Chandler
  Bennett, Mark~A. Berrill, Nick~S. Blunt, Edgar Josu{\'{e}}~Landinez Borda,
  Michele Casula, David~M. Ceperley, Simone Chiesa, Bryan~K. Clark, Raymond~C.
  Clay, Kris~T. Delaney, Mark Dewing, Kenneth~P. Esler, Hongxia Hao, Olle
  Heinonen, Paul R~C Kent, Jaron~T. Krogel, Ilkka Kyl{\"{a}}np{\"{a}}{\"{a}},
  Ying~Wai Li, M.~Graham Lopez, Ye~Luo, Fionn~D. Malone, Richard~M. Martin,
  Amrita Mathuriya, Jeremy McMinis, Cody~A. Melton, Lubos Mitas, Miguel~A.
  Morales, Eric Neuscamman, William~D. Parker, Sergio~D. {Pineda Flores},
  Nichols~A. Romero, Brenda~M. Rubenstein, Jacqueline A~R Shea, Hyeondeok Shin,
  Luke Shulenburger, Andreas~F. Tillack, Joshua~P. Townsend, Norm~M. Tubman,
  Brett {Van Der Goetz}, Jordan~E. Vincent, D~ChangMo Yang, Yubo Yang, Shuai
  Zhang, and Luning Zhao.
\newblock {QMCPACK : an open source ab initio quantum Monte Carlo package for
  the electronic structure of atoms, molecules and solids}.
\newblock \emph{Journal of Physics: Condensed Matter}, 30\penalty0
  (19):\penalty0 195901, 5 2018.
\newblock \doi{10.1088/1361-648X/aab9c3}.
\newblock URL
  \url{https://iopscience.iop.org/article/10.1088/1361-648X/aab9c3}.

\bibitem[Kent et~al.(2020)Kent, Annaberdiyev, Benali, Bennett, {Landinez
  Borda}, Doak, Hao, Jordan, Krogel, Kyl{\"{a}}np{\"{a}}{\"{a}}, Lee, Luo,
  Malone, Melton, Mitas, Morales, Neuscamman, Reboredo, Rubenstein, Saritas,
  Upadhyay, Wang, Zhang, and Zhao]{Kent2020}
P.~R.~C. Kent, Abdulgani Annaberdiyev, Anouar Benali, M.~Chandler Bennett,
  Edgar~Josu{\'{e}} {Landinez Borda}, Peter Doak, Hongxia Hao, Kenneth~D.
  Jordan, Jaron~T. Krogel, Ilkka Kyl{\"{a}}np{\"{a}}{\"{a}}, Joonho Lee,
  Ye~Luo, Fionn~D. Malone, Cody~A. Melton, Lubos Mitas, Miguel~A. Morales, Eric
  Neuscamman, Fernando~A. Reboredo, Brenda Rubenstein, Kayahan Saritas, Shiv
  Upadhyay, Guangming Wang, Shuai Zhang, and Luning Zhao.
\newblock {QMCPACK: Advances in the development, efficiency, and application of
  auxiliary field and real-space variational and diffusion quantum Monte
  Carlo}.
\newblock \emph{The Journal of Chemical Physics}, 152\penalty0 (17):\penalty0
  174105, 5 2020.
\newblock ISSN 0021-9606.
\newblock \doi{10.1063/5.0004860}.
\newblock URL \url{https://doi.org/10.1063/5.0004860
  http://aip.scitation.org/doi/10.1063/5.0004860}.

\bibitem[Pierleoni et~al.(2008)Pierleoni, Delaney, Morales, Ceperley, and
  Holzmann]{pierleoni2008}
Carlo Pierleoni, Kris~T. Delaney, Miguel~A. Morales, David~M. Ceperley, and
  Markus Holzmann.
\newblock Trial wave functions for high-pressure metallic hydrogen.
\newblock \emph{Computer Physics Communications}, 179\penalty0 (1):\penalty0
  89--97, 2008.
\newblock ISSN 0010-4655.
\newblock \doi{https://doi.org/10.1016/j.cpc.2008.01.041}.
\newblock URL
  \url{https://www.sciencedirect.com/science/article/pii/S0010465508000210}.
\newblock Special issue based on the Conference on Computational Physics 2007.

\bibitem[Holzmann et~al.(2016)Holzmann, Clay~III, Morales, Tubman, Ceperley,
  and Pierleoni]{holzmann2016theory}
Markus Holzmann, Raymond~C Clay~III, Miguel~A Morales, Norm~M Tubman, David~M
  Ceperley, and Carlo Pierleoni.
\newblock Theory of finite size effects for electronic quantum monte carlo
  calculations of liquids and solids.
\newblock \emph{Physical Review B}, 94\penalty0 (3):\penalty0 035126, 2016.

\bibitem[Chiesa et~al.(2005)Chiesa, Ceperley, and Zhang]{Chiesa2005}
Simone Chiesa, D.~M. Ceperley, and Shiwei Zhang.
\newblock {Accurate, Efficient, and Simple Forces Computed with Quantum Monte
  Carlo Methods}.
\newblock \emph{Phys. Rev. Lett.}, 94\penalty0 (3):\penalty0 036404, 1 2005.
\newblock ISSN 0031-9007.
\newblock \doi{10.1103/PhysRevLett.94.036404}.
\newblock URL \url{https://link.aps.org/doi/10.1103/PhysRevLett.94.036404}.

\bibitem[Thompson et~al.(2022)Thompson, Aktulga, Berger, Bolintineanu, Brown,
  Crozier, in~'t Veld, Kohlmeyer, Moore, Nguyen, Shan, Stevens, Tranchida,
  Trott, and Plimpton]{LAMMPS}
A.~P. Thompson, H.~M. Aktulga, R.~Berger, D.~S. Bolintineanu, W.~M. Brown,
  P.~S. Crozier, P.~J. in~'t Veld, A.~Kohlmeyer, S.~G. Moore, T.~D. Nguyen,
  R.~Shan, M.~J. Stevens, J.~Tranchida, C.~Trott, and S.~J. Plimpton.
\newblock {LAMMPS} - a flexible simulation tool for particle-based materials
  modeling at the atomic, meso, and continuum scales.
\newblock \emph{Comp. Phys. Comm.}, 271:\penalty0 108171, 2022.
\newblock \doi{10.1016/j.cpc.2021.108171}.

\bibitem[Li et~al.(2026)Li, Gomez, Cai, Zhang, Fu, Jia, Feldman, Blumer, Higer,
  Hirshberg, et~al.]{li2026fix}
Yifan Li, Axel Gomez, Kehan Cai, Chunyi Zhang, Li~Fu, Weile Jia, Yotam~MY
  Feldman, Ofir Blumer, Jacob Higer, Barak Hirshberg, et~al.
\newblock fix pimd/langevin: An efficient implementation of path integral
  molecular dynamics in lammps.
\newblock \emph{The Journal of Chemical Physics}, 164\penalty0 (14), 2026.

\bibitem[Litman et~al.(2024)Litman, Kapil, Feldman, Tisi, Begušić, Fidanyan,
  Fraux, Higer, Kellner, Li, Pós, Stocco, Trenins, Hirshberg, Rossi, and
  Ceriotti]{iPI3}
Yair Litman, Venkat Kapil, Yotam M.~Y. Feldman, Davide Tisi, Tomislav
  Begušić, Karen Fidanyan, Guillaume Fraux, Jacob Higer, Matthias Kellner,
  Tao~E. Li, Eszter~S. Pós, Elia Stocco, George Trenins, Barak Hirshberg,
  Mariana Rossi, and Michele Ceriotti.
\newblock i-pi 3.0: A flexible and efficient framework for advanced atomistic
  simulations.
\newblock \emph{The Journal of Chemical Physics}, 161\penalty0 (6):\penalty0
  062504, 08 2024.
\newblock ISSN 0021-9606.
\newblock \doi{10.1063/5.0215869}.
\newblock URL \url{https://doi.org/10.1063/5.0215869}.

\bibitem[Wigner(1932)]{Wigner1932}
E.~Wigner.
\newblock On the quantum correction for thermodynamic equilibrium.
\newblock \emph{Phys. Rev.}, 40:\penalty0 749--759, 6 1932.
\newblock \doi{10.1103/PhysRev.40.749}.
\newblock URL \url{https://link.aps.org/doi/10.1103/PhysRev.40.749}.

\bibitem[Batatia et~al.(2022)Batatia, Kovacs, Simm, Ortner, and
  Cs{\'a}nyi]{batatia2022mace}
Ilyes Batatia, David~P Kovacs, Gregor Simm, Christoph Ortner, and G{\'a}bor
  Cs{\'a}nyi.
\newblock Mace: Higher order equivariant message passing neural networks for
  fast and accurate force fields.
\newblock \emph{Advances in Neural Information Processing Systems},
  35:\penalty0 11423--11436, 2022.

\bibitem[Zhang et~al.(2026)Zhang, Peng, Cai, Li, Zhou, Zeng, Guo, Zhang, Li,
  Jiang, Zhu, Jia, Zhang, and Wang]{zhang2026graphneuralnetworkera}
Duo Zhang, Anyang Peng, Chun Cai, Wentao Li, Yuanchang Zhou, Jinzhe Zeng,
  Mingyu Guo, Chengqian Zhang, Bowen Li, Hong Jiang, Tong Zhu, Weile Jia,
  Linfeng Zhang, and Han Wang.
\newblock A graph neural network for the era of large atomistic models.
\newblock \emph{arXiv preprint arXiv:2506.01686}, 2026.
\newblock URL \url{https://arxiv.org/abs/2506.01686}.

\bibitem[Peng et~al.(2026)Peng, Cai, Guo, Zhang, Zhang, Jiang, Wang, Loew, Wu,
  E, et~al.]{peng2026lambench}
Anyang Peng, Chun Cai, Mingyu Guo, Duo Zhang, Chengqian Zhang, Wanrun Jiang,
  Yinan Wang, Antoine Loew, Chengkun Wu, Weinan E, et~al.
\newblock Lambench: a benchmark for large atomistic models.
\newblock \emph{npj Computational Materials}, 2026.

\bibitem[Kovács et~al.(2023)Kovács, Batatia, Arany, and
  Csányi]{kovacs_evaluation_2023}
Dávid~Péter Kovács, Ilyes Batatia, Eszter~Sára Arany, and Gábor Csányi.
\newblock Evaluation of the {MACE} force field architecture: {From} medicinal
  chemistry to materials science.
\newblock \emph{The Journal of Chemical Physics}, 159\penalty0 (4):\penalty0
  044118, July 2023.
\newblock ISSN 0021-9606, 1089-7690.
\newblock \doi{10.1063/5.0155322}.
\newblock URL
  \url{https://pubs.aip.org/jcp/article/159/4/044118/2904837/Evaluation-of-the-MACE-force-field-architecture}.

\bibitem[Zhang et~al.(2025)Zhang, Peng, Cai, Li, Zhou, Zeng, Guo, Zhang, Li,
  Jiang, et~al.]{zhang2025graph}
Duo Zhang, Anyang Peng, Chun Cai, Wentao Li, Yuanchang Zhou, Jinzhe Zeng,
  Mingyu Guo, Chengqian Zhang, Bowen Li, Hong Jiang, et~al.
\newblock Graph neural network model for the era of large atomistic models.
\newblock \emph{arXiv preprint arXiv:2506.01686}, 2025.

\bibitem[Pierleoni and Ceperley(2006)]{pierleoni2006coupled}
Carlo Pierleoni and David~M Ceperley.
\newblock The coupled electron-ion monte carlo method.
\newblock \emph{Computer Simulations in Condensed Matter Systems: From
  Materials to Chemical Biology Volume 1}, pages 641--683, 2006.

\bibitem[Pierleoni et~al.(2016)Pierleoni, Morales, Rillo, Holzmann, and
  Ceperley]{pierleoni2016liquid}
Carlo Pierleoni, Miguel~A Morales, Giovanni Rillo, Markus Holzmann, and David~M
  Ceperley.
\newblock Liquid--liquid phase transition in hydrogen by coupled electron--ion
  monte carlo simulations.
\newblock \emph{Proceedings of the National Academy of Sciences}, 113\penalty0
  (18):\penalty0 4953--4957, 2016.

\bibitem[Pierleoni et~al.(2018)Pierleoni, Holzmann, and
  Ceperley]{pierleoni2018CPP}
Carlo Pierleoni, Markus Holzmann, and David~M. Ceperley.
\newblock Local structure in dense hydrogen at the liquid–liquid phase
  transition by coupled electron–ion monte carlo.
\newblock \emph{Contributions to Plasma Physics}, 58\penalty0 (2-3):\penalty0
  99--106, 2018.
\newblock \doi{https://doi.org/10.1002/ctpp.201700184}.
\newblock URL
  \url{https://onlinelibrary.wiley.com/doi/abs/10.1002/ctpp.201700184}.

\bibitem[Morales et~al.(2010)Morales, Pierleoni, Schwegler, and
  Ceperley]{morales2010evidence}
Miguel~A Morales, Carlo Pierleoni, Eric Schwegler, and David~M Ceperley.
\newblock Evidence for a first-order liquid-liquid transition in high-pressure
  hydrogen from ab initio simulations.
\newblock \emph{Proceedings of the National Academy of Sciences}, 107\penalty0
  (29):\penalty0 12799--12803, 2010.

\bibitem[Tian et~al.(2019)Tian, Liu, Yuan, Chen, and Kuan]{tian2019first}
Chunling Tian, Fusheng Liu, Hongkuan Yuan, Hong Chen, and Anlong Kuan.
\newblock First-order liquid-liquid phase transition in compressed hydrogen and
  critical point.
\newblock \emph{The Journal of Chemical Physics}, 150\penalty0 (20), 2019.

\bibitem[Khrapak and Yurchenko(2021)]{khrapak2021}
Sergey~A Khrapak and Stanislav~O Yurchenko.
\newblock Entropy of simple fluids with repulsive interactions near freezing.
\newblock \emph{The Journal of Chemical Physics}, 155\penalty0 (13), 2021.

\bibitem[Nicholson et~al.(2021)Nicholson, Gao, McDonnell, Sluss, and
  Keffer]{nicholson2021}
Donald~M Nicholson, CY~Gao, Marshall~T McDonnell, Clifton~C Sluss, and David~J
  Keffer.
\newblock Entropy pair functional theory: Direct entropy evaluation spanning
  phase transitions.
\newblock \emph{Entropy}, 23\penalty0 (2):\penalty0 234, 2021.

\bibitem[Gartner~III et~al.(2022)Gartner~III, Piaggi, Car, Panagiotopoulos, and
  Debenedetti]{gartner2022liquid}
Thomas~E Gartner~III, Pablo~M Piaggi, Roberto Car, Athanassios~Z
  Panagiotopoulos, and Pablo~G Debenedetti.
\newblock Liquid-liquid transition in water from first principles.
\newblock \emph{Physical review letters}, 129\penalty0 (25):\penalty0 255702,
  2022.

\bibitem[Nikitiuk et~al.(2022)Nikitiuk, Salikova, Kondratyuk, and
  Pisarev]{nikitiuk2022pair}
BI~Nikitiuk, DI~Salikova, ND~Kondratyuk, and VV~Pisarev.
\newblock Pair entropy and universal viscosity scaling for molecular systems
  via molecular dynamics simulations.
\newblock \emph{Journal of Molecular Liquids}, 368:\penalty0 120714, 2022.

\bibitem[Lukyanchuk et~al.(2026)Lukyanchuk, Norman, Kondratyuk, and
  Saitov]{lukyanchuk2026influence}
Vyacheslav~I Lukyanchuk, Genri~E Norman, Nikolay~D Kondratyuk, and IInur~M
  Saitov.
\newblock Influence of nuclear quantum effects on metastable states of warm
  dense hydrogen.
\newblock \emph{Scientific Reports}, 16:\penalty0 16435, 2026.

\bibitem[Binder(1981)]{binder_finite_nodate}
K~Binder.
\newblock Finite size scaling analysis of ising model block distribution
  functions.
\newblock \emph{Zeitschrift fur Physik B Condensed Matter}, 43\penalty0
  (2):\penalty0 119--140, 1981.

\bibitem[Wilding(1997)]{wilding_simulation_1997}
Nigel~B Wilding.
\newblock Simulation studies of fluid critical behaviour.
\newblock \emph{Journal of Physics: Condensed Matter}, 9\penalty0 (3):\penalty0
  585--612, 1 1997.
\newblock ISSN 0953-8984, 1361-648X.
\newblock \doi{10.1088/0953-8984/9/3/002}.
\newblock URL
  \url{https://iopscience.iop.org/article/10.1088/0953-8984/9/3/002}.

\bibitem[Binder et~al.(2019)Binder, Heermann, and
  Binder]{binder_book_monte_carlo}
Kurt Binder, Dieter~W Heermann, and K~Binder.
\newblock \emph{Monte Carlo simulation in statistical physics}, volume~8.
\newblock Springer, 2019.

\bibitem[Ferrenberg and Swendsen(1989)]{ferrenberg1989optimized}
Alan~M Ferrenberg and Robert~H Swendsen.
\newblock Optimized monte carlo data analysis.
\newblock \emph{Physical review letters}, 63\penalty0 (12):\penalty0 1195,
  1989.

\bibitem[Newman and Barkema(1999)]{newman1999monte}
Mark Newman and Gerard Barkema.
\newblock Monte carlo methods in statistical physics chapter 1-4.
\newblock \emph{New York, USA}, 1999.

\bibitem[Ceperley(1995)]{ceperley1995path}
David~M Ceperley.
\newblock Path integrals in the theory of condensed helium.
\newblock \emph{Reviews of Modern Physics}, 67\penalty0 (2):\penalty0 279,
  1995.

\bibitem[Scandolo(2003)]{scandolo2003liquid}
Sandro Scandolo.
\newblock Liquid--liquid phase transition in compressed hydrogen from
  first-principles simulations.
\newblock \emph{Proceedings of the National Academy of Sciences}, 100\penalty0
  (6):\penalty0 3051--3053, 2003.

\bibitem[Bonev et~al.(2004)Bonev, Militzer, and Galli]{Bonev2004}
Stanimir~A. Bonev, Burkhard Militzer, and Giulia Galli.
\newblock Ab initio simulations of dense liquid deuterium: Comparison with
  gas-gun shock-wave experiments.
\newblock \emph{Phys. Rev. B}, 69:\penalty0 014101, Jan 2004.
\newblock \doi{10.1103/PhysRevB.69.014101}.
\newblock URL \url{https://link.aps.org/doi/10.1103/PhysRevB.69.014101}.

\bibitem[Tamblyn and Bonev(2010)]{Tamblyn2010}
Isaac Tamblyn and Stanimir~A. Bonev.
\newblock Structure and phase boundaries of compressed liquid hydrogen.
\newblock \emph{Phys. Rev. Lett.}, 104:\penalty0 065702, Feb 2010.
\newblock \doi{10.1103/PhysRevLett.104.065702}.
\newblock URL \url{https://link.aps.org/doi/10.1103/PhysRevLett.104.065702}.

\bibitem[Lorenzen et~al.(2010)Lorenzen, Holst, and Redmer]{lorenzen2010first}
Winfried Lorenzen, Bastian Holst, and Ronald Redmer.
\newblock First-order liquid-liquid phase transition in dense hydrogen.
\newblock \emph{Physical Review B—Condensed Matter and Materials Physics},
  82\penalty0 (19):\penalty0 195107, 2010.

\bibitem[Karasiev et~al.(2021)Karasiev, Hinz, Hu, and
  Trickey]{karasiev2021liquid}
Valentin~V Karasiev, Joshua Hinz, SX~Hu, and Samuel~B Trickey.
\newblock On the liquid--liquid phase transition of dense hydrogen.
\newblock \emph{Nature}, 600\penalty0 (7889):\penalty0 E12--E14, 2021.

\bibitem[Istas et~al.(2025)Istas, Jensen, Yang, Holzmann, Pierleoni, and
  Ceperley]{istas2025liquid}
Mathieu Istas, Scott Jensen, Yubo Yang, Markus Holzmann, Carlo Pierleoni, and
  David~M Ceperley.
\newblock Liquid-liquid phase transition of hydrogen and its critical point:
  Analysis from ab initio simulation and a machine-learned potential.
\newblock \emph{Physical Review E}, 111\penalty0 (4):\penalty0 045307, 2025.

\bibitem[Tenti et~al.(2025)Tenti, Jäckl, Nakano, Rupp, and
  Casula]{tenti_hydrogen_2025}
Giacomo Tenti, Bastian Jäckl, Kousuke Nakano, Matthias Rupp, and Michele
  Casula.
\newblock Hydrogen liquid-liquid transition from first principles and machine
  learning.
\newblock \emph{Physical Review B}, 112\penalty0 (10):\penalty0 104208,
  September 2025.
\newblock ISSN 2469-9950, 2469-9969.
\newblock \doi{10.1103/pbrk-3zgd}.
\newblock URL \url{https://link.aps.org/doi/10.1103/pbrk-3zgd}.

\bibitem[Morales et~al.(2013)Morales, McMahon, Pierleoni, and
  Ceperley]{morales2013}
Miguel~A. Morales, Jeffrey~M. McMahon, Carlo Pierleoni, and David~M. Ceperley.
\newblock Nuclear quantum effects and nonlocal exchange-correlation functionals
  applied to liquid hydrogen at high pressure.
\newblock \emph{Phys. Rev. Lett.}, 110:\penalty0 065702, Feb 2013.
\newblock \doi{10.1103/PhysRevLett.110.065702}.
\newblock URL \url{https://link.aps.org/doi/10.1103/PhysRevLett.110.065702}.

\bibitem[Geng et~al.(2019)Geng, Wu, Marqu{\'e}s, and
  Ackland]{geng2019thermodynamic}
Hua~Y Geng, Q~Wu, Miriam Marqu{\'e}s, and Graeme~J Ackland.
\newblock Thermodynamic anomalies and three distinct liquid-liquid transitions
  in warm dense liquid hydrogen.
\newblock \emph{Physical Review B}, 100\penalty0 (13):\penalty0 134109, 2019.

\bibitem[Hinz et~al.(2020)Hinz, Karasiev, Hu, Zaghoo,
  Mej\'{\i}a-Rodr\'{\i}guez, Trickey, and Calder\'{\i}n]{Hinz2020}
Joshua Hinz, Valentin~V. Karasiev, S.~X. Hu, Mohamed Zaghoo, Daniel
  Mej\'{\i}a-Rodr\'{\i}guez, S.~B. Trickey, and L.~Calder\'{\i}n.
\newblock Fully consistent density functional theory determination of the
  insulator-metal transition boundary in warm dense hydrogen.
\newblock \emph{Phys. Rev. Res.}, 2:\penalty0 032065(R), Sep 2020.
\newblock \doi{10.1103/PhysRevResearch.2.032065}.
\newblock URL \url{https://link.aps.org/doi/10.1103/PhysRevResearch.2.032065}.

\bibitem[van~de Bund et~al.(2021)van~de Bund, Wiebe, and Ackland]{Bund2021}
Sebastiaan van~de Bund, Heather Wiebe, and Graeme~J. Ackland.
\newblock Isotope quantum effects in the metallization transition in liquid
  hydrogen.
\newblock \emph{Phys. Rev. Lett.}, 126:\penalty0 225701, Jun 2021.
\newblock \doi{10.1103/PhysRevLett.126.225701}.

\bibitem[Bergermann et~al.(2024)Bergermann, Kleindienst, and
  Redmer]{bergermann2024}
Armin Bergermann, Lucas Kleindienst, and Ronald Redmer.
\newblock Nonmetal-to-metal transition in liquid hydrogen using density
  functional theory and the heyd–scuseria–ernzerhof exchange-correlation
  functional.
\newblock \emph{The Journal of Chemical Physics}, 161\penalty0 (23):\penalty0
  234303, 12 2024.
\newblock ISSN 0021-9606.
\newblock \doi{10.1063/5.0241111}.
\newblock URL \url{https://doi.org/10.1063/5.0241111}.

\bibitem[Mazzola et~al.(2018)Mazzola, Helled, and Sorella]{Mazzola2018}
Guglielmo Mazzola, Ravit Helled, and Sandro Sorella.
\newblock {Phase Diagram of Hydrogen and a Hydrogen-Helium Mixture at Planetary
  Conditions by Quantum Monte Carlo Simulations}.
\newblock \emph{Physical Review Letters}, 120:\penalty0 025701, 2018.
\newblock \doi{10.1103/PhysRevLett.120.025701}.
\newblock URL
  \url{https://journals.aps.org/prl/pdf/10.1103/PhysRevLett.120.025701}.

\bibitem[Tirelli et~al.(2022)Tirelli, Tenti, Nakano, and
  Sorella]{tirelli2022high}
Andrea Tirelli, Giacomo Tenti, Kousuke Nakano, and Sandro Sorella.
\newblock High-pressure hydrogen by machine learning and quantum monte carlo.
\newblock \emph{Physical Review B}, 106\penalty0 (4):\penalty0 L041105, 2022.

\bibitem[Zha et~al.(2017)Zha, Liu, Tse, and Hemley]{Zha2017}
Chang-Sheng Zha, Hanyu Liu, John~S Tse, and Russell~J Hemley.
\newblock {Melting and High P-T Transitions of Hydrogen up to 300 GPa}.
\newblock \emph{Phys. Rev. Letts.}, 119:\penalty0 075302, 2017.
\newblock \doi{10.1103/PhysRevLett.119.075302}.
\newblock URL
  \url{https://journals.aps.org/prl/pdf/10.1103/PhysRevLett.119.075302}.

\bibitem[Luo et~al.(2026)Luo, Ichibha, Ma, Liu, Li, Maezono, Zhong, Xie, Liu,
  Hemley, and Ma]{Luo2026}
Jie Luo, Tom Ichibha, Yuan Ma, Yuan Liu, Jiaxiang Li, Ryo Maezono, Xin Zhong,
  Yu~Xie, Hanyu Liu, Russell~J. Hemley, and Yanming Ma.
\newblock Quantum melting behavior of dense solid hydrogen from large-scale
  machine-learned potential simulations.
\newblock \emph{Phys. Rev. B}, 113:\penalty0 184116, May 2026.
\newblock \doi{10.1103/47cf-cj2s}.
\newblock URL \url{https://link.aps.org/doi/10.1103/47cf-cj2s}.

\end{thebibliography}
\end{document}